\documentclass{layoutArxiv/egpubl}
\usepackage{layoutArxiv/hpg21}
 
\SpecialIssuePaper         % uncomment for final version of Computer Graphics Forum, special issue
\usepackage[T1]{fontenc}
\usepackage{"./layoutArxiv/dfadobe"}  

\usepackage[export]{adjustbox}
\usepackage{graphicx}
\usepackage{color}
\usepackage{xcolor}
\usepackage{subcaption}
\usepackage{float}
\usepackage{tabu}
\usepackage{booktabs}
\usepackage{multirow}
\usepackage{amsmath}
\usepackage{listings}[language=C++]
\usepackage{hyperref}
\usepackage{overpic}
\usepackage[framemethod=tikz]{mdframed}
\usepackage{lipsum}
\usetikzlibrary{shadows}

\newmdenv[linewidth= 1pt,linecolor= white, tikzsetting={draw=black, line width = 2pt, dashed,%
dash pattern = on 4pt off 3pt}]{myshadowbox}
                
\usepackage{todonotes}

\usepackage{textcomp}
\usepackage{colortbl}
\usepackage{soul}

\soulregister\ref7
\soulregister\cite7

\newcommand*\rot{\rotatebox{90}}

\definecolor{codegreen}{rgb}{0,0.6,0}
\definecolor{codegray}{rgb}{0.5,0.5,0.5}
\definecolor{codepurple}{rgb}{0.58,0,0.82}
\definecolor{backcolour}{rgb}{0.95,0.95,0.92}

\lstdefinestyle{mystyle}{ 
    commentstyle=\color{codegreen},
    keywordstyle=\color{magenta},
    numberstyle=\tiny\color{codegray},
    stringstyle=\color{codepurple},
    basicstyle=\ttfamily\small,
    breakatwhitespace=false,         
    breaklines=true,                 
    captionpos=b,                    
    keepspaces=true,                 
    numbers=left,                    
    showspaces=false,                
    showstringspaces=false,
    showtabs=false,                  
    tabsize=2
}
\title{GPU-Accelerated LOD Generation for Point Clouds}

\author[M. Schütz \& B. Kerbl \& P. Klaus \& M. Wimmer]
{\parbox{\textwidth}{
    \centering Markus Schütz$^{1}$, Bernhard Kerbl$^{3}$, Philip Klaus$^{2}$, Michael Wimmer$^{1}$
        }
        \\
{\parbox{\textwidth}{\centering $^1$TU Wien, 
         $^2$Austrian Institute of Technology, 
         $^3$Inria, Université Côte d'Azur
       }
}
}

\biberVersion
\BibtexOrBiblatex
\usepackage[backend=biber,bibstyle=./layoutArxiv/EG,citestyle=alphabetic,backref=true]{biblatex} 
\begin{document}

\teaser{
    
    \begin{overpic}[abs,unit=1mm,scale=.25]{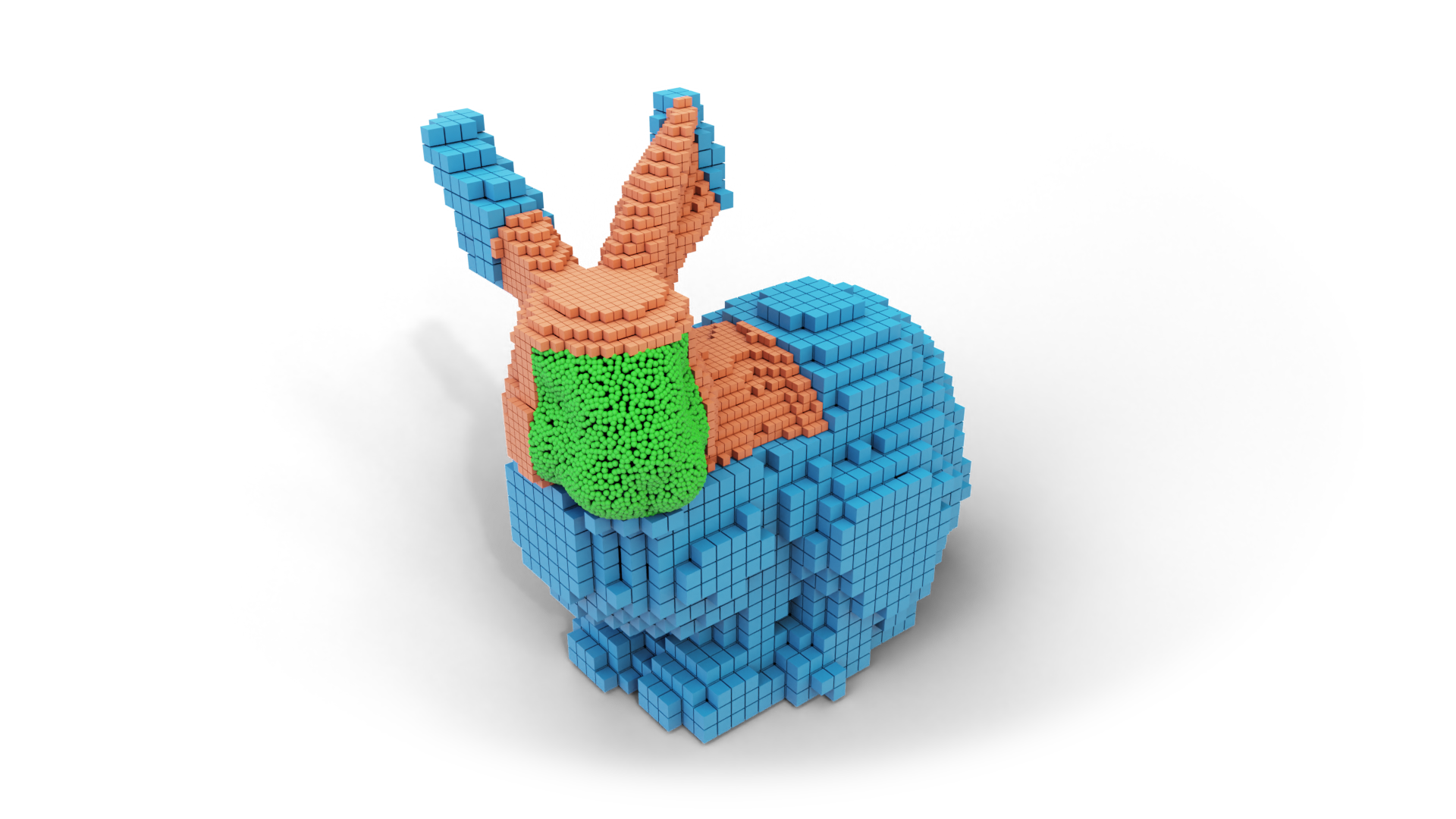}
    \put(1,70){
    \begin{minipage}{14em}    
    \begin{myshadowbox}
    \includegraphics[width=3cm,trim={18cm 3cm 22cm 6cm}]{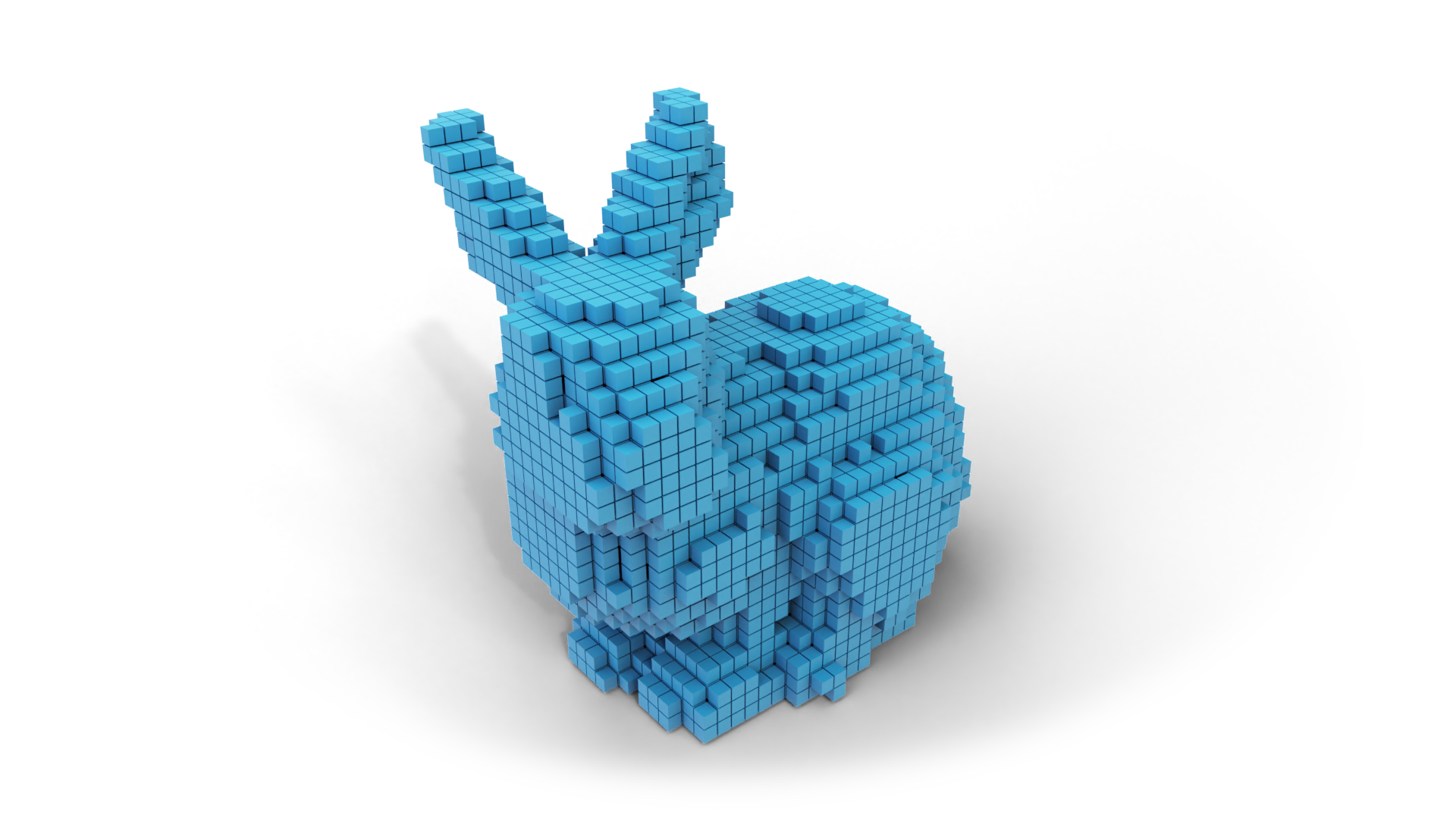}\\
    {\small
    At the coarsest level, we use $128^3$ (shown: $32^3$) surface-voxels to represent the model.}
    \end{myshadowbox}
    \end{minipage}
    }
    \put(120,70){
    \begin{minipage}{14em}    
    \begin{myshadowbox}
    \includegraphics[width=3cm,trim={18cm 3cm 22cm 6cm}]{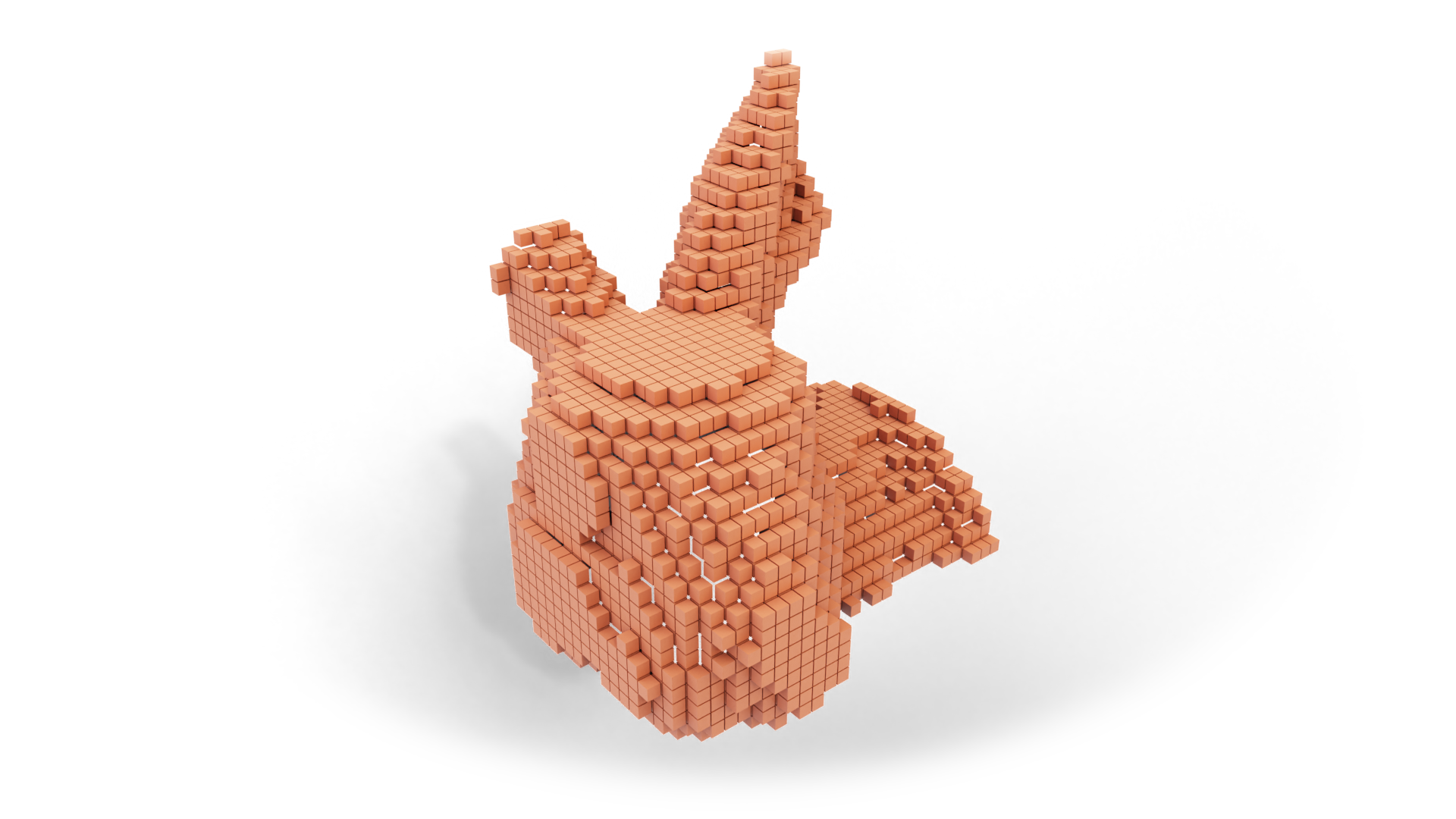}\\
    {\small
    Intermediate nodes add further detail by replacing one octant ($64^3$ voxels) of their parent node with a finer, higher-resolution representation ($128^3$ voxels).}
    \end{myshadowbox}
    \end{minipage}
    }
    \put(1,20){
    \begin{minipage}{14em}    
    \begin{myshadowbox}
    \vspace{0.2cm}
    \includegraphics[width=3cm,trim={15cm 7cm 25cm 10cm}]{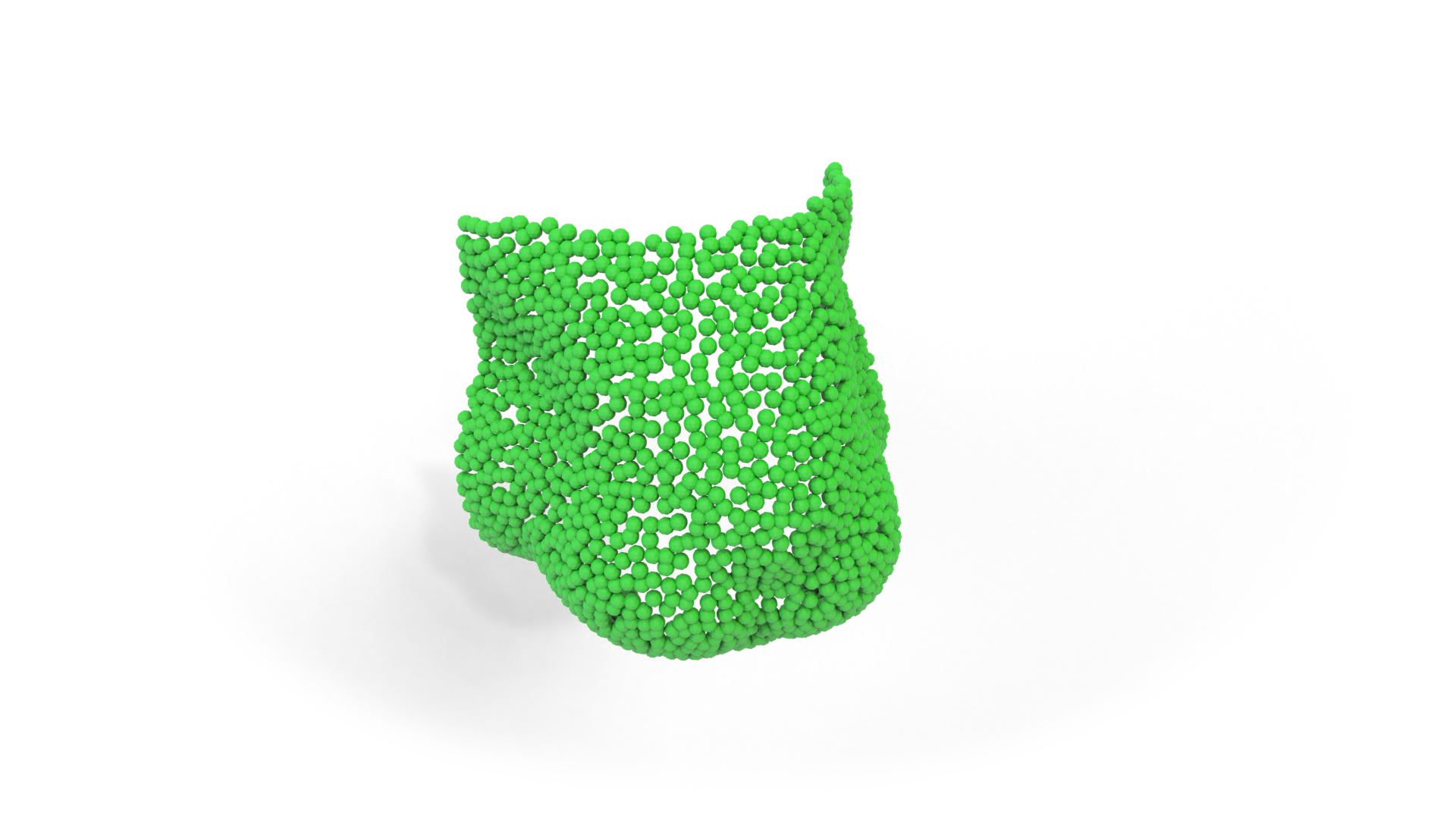}\\
    {\small
    Leaf nodes model the original, full-precision point cloud data.}
    \end{myshadowbox}
    \end{minipage}
    }
    \end{overpic}
    \centering
    \caption{The data structure is a hybrid voxel-point octree that utilizes color-filtered voxels for lower LODs, but displays the original point data at the highest LOD. As most grid cells of each node are empty, surface voxels are stored as vertices in a node's vertex buffer. During rendering, we aim to rasterize pixel-sized voxels to give viewers the impression that they are looking at full-precision point cloud data. }
    \label{fig:teaser}
}

\maketitle

\begin{abstract}

\textbf{About:}
We introduce a GPU-accelerated LOD construction process that creates a hybrid voxel-point-based variation of the widely used layered point cloud (LPC) structure for LOD rendering and streaming. The massive performance improvements provided by the GPU allow us to improve the quality of lower LODs via color filtering while still increasing construction speed compared to the non-filtered, CPU-based state of the art.

\vspace*{3px}
\noindent
\textbf{Background:} LOD structures are required to render hundreds of millions to trillions of points, but constructing them takes time.

\vspace*{3px}
\noindent
\textbf{Results:} LOD structures suitable for rendering and streaming are constructed at rates of about 1 billion points per second (with color filtering) to 4 billion points per second (sample-picking/random sampling, state of the art) on an RTX 3090 -- an improvement of a factor of 80 to 400 times over the CPU-based state of the art (12 million points per second). Due to being in-core, model sizes are limited to about 500 million points per 24GB memory.

\vspace*{3px}
\noindent
\textbf{Discussion:} Our method currently focuses on maximizing in-core construction speed on the GPU. Issues such as out-of-core construction of arbitrarily large data sets are not addressed, but we expect it to be suitable as a component of bottom-up out-of-core LOD construction schemes.

\end{abstract}

%\bibliographystyle{alpha}

% \clearpage

\section{Introduction}

Point clouds -- 3D models made of colored vertices -- are widely used in the geospatial industry, archaeology, construction and other fields that operate on digital twins of real environments and objects. 3D scanning devices such as laser scanners digitize real surfaces in a point-wise fashion, resulting in hundreds of millions to trillions of points~\cite{AHN2,USGS:3DEP,USGS:Entwine}. 

Processing and visualizing these massive amounts of points is an ongoing field of research, where advances in rendering hardware and algorithms compete with the ever growing size of the data sets. Compared to triangle meshes, point clouds tend to be significantly larger. While a few textured triangles are sufficient to give the illusion of a detailed and closed surface, thousands to millions of vertex-colored point samples are necessary to represent the same surface with a similar amount of detail and without holes between points upon closer inspection. Although point clouds can, and frequently are, converted to textured meshes, there are use cases and processes that favour the original point cloud data. For example, we may want to produce various different products out of a point cloud (digital elevation models (DEM), digital surface models (DSM), meshes), and we may want to recreate these from scratch with different settings (resolution, quality) or apply improved meshing algorithms in the future. We may also want to measure directly on the original, full-precision scan data, rather than derivatives with potentially reduced or made-up information. Furthermore, creating meshes is a time-consuming and still error-prone process, so operating directly on the source data may lead to faster and more accurate measurements. 

The visualization of hundreds of millions of points requires level-of-detail structures that allow the rendering engine to load and display only the smallest subset of data with the biggest impact on visual quality. Important characteristics of LOD structures are a reduction in load times, memory usage, and improved rendering performance as only a fraction of the data is loaded and processed at any given time. LOD structures may also improve the visual quality via precomputed antialiasing (color filtering), similar to mip mapping and virtual texturing. 

Layered point clouds (LPC)~\cite{GOBBETTI2004} and its variations are one of the most widely used LOD structures for point clouds. LPCs were originally introduced as a binary tree where each node contains a random subset of the point cloud. Lower-level nodes contain coarse subsets of large areas, while higher-level nodes contain increasingly more detailed subsets of smaller areas. During rendering, we draw nodes from coarse to fine until we reach the desired level of detail. More detailed nodes are discarded, as well as all the nodes outside the view frustum. 

One of the main issues of LOD structures is that creating them is computationally expensive and takes time. The fastest LPC generation methods have a throughput of about 2.5 to 11 million points per second~\cite{Kang2019, Bormann:PCI, 9671659, SCHUETZ-2020-MPC}, which makes the inspection of typical data sets with hundreds of millions of points tedious and time-intensive.

In this paper, we propose a GPU-accelerated LOD-generation process that creates a hybrid voxel-point-based variation of layered point clouds, using an octree. Our method constructs the octree up to two orders of magnitude faster than the current state of the art and generates color-filtered voxels in lower LODs to improve quality. We propose to use voxels instead of points in lower LODs because quantized integer voxel coordinates offer higher compression rates compared to full-precision point coordinates, and thus reduce transfer times when streaming lower LODs in web browsers. It should be noted, however, that our process is easily modified to support points in lower LODs, if desired.

Our contributions to the state of the art are: 

\begin{itemize}
    \item A hybrid voxel-point-based variation of Layered Point Clouds~\cite{GOBBETTI2004} and Modifiable Nested Octree~\cite{scheiblauer2011}) that incorporates color-filtering and enables more efficient streaming of lower LODs due to higher compression rates of quantized voxel coordinates. It can simultaneously also be seen as a point-based variation of the hybrid voxel-triangle structure proposed by Chajdas et al.~\cite{Chajdas2014Scalable}.
    \item A CUDA-accelerated variation of a fast bottom-up LOD construction scheme for point clouds~\cite{SCHUETZ-2020-MPC} that is up to two orders of magnitudes faster than its CPU-based predecessor. 
\end{itemize}

%\clearpage
\section{Related Work}

Level-of-Detail methods allow us to render large 3D models that would otherwise be slow to render and/or too large to fit in memory~\cite{Luebke:2002:LDG:2821571}. LOD methods may offer multiple versions of a model with varying resolutions, spatially partition them so that we can load and render only subsets that are within the view-frustum (view-dependent LOD), and precompute anti-aliased geometry and textures to create higher-quality images with lower processing effort (e.g., mip mapping, virtual texturing). Of special note is Nanite~\cite{karis2021nanite}, which recently made it possible to efficiently stream and render massive, complex triangle meshes in a game engine. 

\subsection{Point-Based Level-of-Detail Structures}

QSplat~\cite{QSplat} allows displaying large meshes in real-time via point-based rendering of a bounding-sphere hierarchy. The bounding-sphere hierarchy is traversed until a sphere is encountered that is considered to be detailed enough (e.g. a few pixels large), and then rendered as splat with a matching size. Sequential point trees (SPT)~\cite{Dachsbacher2003} is a similar but more GPU-friendly structure that uses a sequentialized version of a bounding sphere hierarchy. The idea is to store the bounding spheres inside an array, sorted by their level of detail. During rendering, we invoke a draw call for the first $X$ elements of the array, depending on the desired level of detail, and in the vertex shader discard the lower LODs while passing the higher, desired LODs to the rasterizer. 

Layered Point Clouds (LPC)~\cite{GOBBETTI2004} was the first GPU-friendly as well as view-dependant LOD structure. Being view-dependant is particularly important as it allows us to adjust the level of detail based on camera position, view direction, and distance to the camera. Distant parts of the model are rendered at lower levels of detail, and regions outside the view frustum are culled entirely. LPCs achieve this by distributing points into a binary tree in which each node contains a random subsample of the original point cloud. Lower levels contain coarser subsamples of the whole data set, while higher levels contain increasingly more detailed subsamples of smaller regions. 

Since then, variations of LPC proposed improvements to various aspects, such as using different tree structures~\cite{InstantPoints, Wand2008, Goswami2010, scheiblauer2011, scheiblauer-thesis, Richter2015, NestedIndexing}, faster LOD construction processes~\cite{scheiblauer-thesis, MartinezRubi2015, Kang2019, SCHUETZ-2020-MPC, Bormann:PCI, 9671659}, sampling strategies that affect construction performance and visual quality~\cite{schuetz-2019-CLOD, Kang2019, SCHUETZ-2020-MPC, VANOOSTEROM2022119}, and editing operations such as efficient selection and deletion of points~\cite{Wand2008, scheiblauer2011}. 

Peak LOD construction performances in terms of throughput were reported to be 2.5~\cite{Kang2019}, 2.6~\cite{Bormann:PCI}, and 11.1 million~\cite{SCHUETZ-2020-MPC} points per second, with the former being an in-core procedure, and the latter two being out-of-core approaches. 

This paper largely adapts the CPU-based LOD construction schemes of Schütz et al.~\cite{SCHUETZ-2020-MPC} into a GPU-based approach, and further extents it by color-filtered lower LODs. Color-filtered lower LODs were already suggested by Wand et al.~\cite{Wand2008} in 2008, but to our knowledge we are the first to attempt to construct layered point clouds with color filtering, while simultaneously improving construction times by two orders of magnitude despite the additional computational effort.

In contrast to geometric LOD streaming and rendering, recent work also explores the use of virtual texturing in order to load and utilize only a small subset of a potentially massive amount of texture data~\cite{HighFidelityPointBased}, potentially with splat-friendly compression~\cite{10.2312:hpg.20211284}. 

\subsection{Voxelization and Voxel-Based Level-of-Detail Structures}

In this paper we focus on surface voxels, i.e., models where each voxel is part of a 3D model's surface. In contrast, solid or volumetric voxel models also specify whether voxels are inside/outside of a 3D model (e.g., Minecraft and similar games), or the density or similar property of a volume at a given position (e.g., CT and MRI data).

\textbf{Voxelization}: Fang and Chen propose a voxelization approach that utilizes the hardware rasterizer to render slices of a 3D model. The pixels of each rendered slice then represent the voxels of the corresponding volume~\cite{HardwareVoxelization}. While efficient, the rasterizer may overlook parts of the surface that intersect pixels but not the sample point within a pixel. Schwarz and Seidel demonstrate a CUDA-based approach for the efficient and conservative or 6-separating surface voxelization of triangles on a $1024^3$ sampling grid~\cite{SurfaceVoxelization}. Conservative meaning that all voxels that are intersected by a triangle are identified, and 6-separating meaning that a smaller amount of intersected voxels that suffices for a grap-free voxelization is identified. (Note: graphics APIs nowadays also support conservative rasterization, potentially resolving limitations that ~\cite{HardwareVoxelization} had back then). 

\textbf{Ray-based LOD}: Sparse-Voxel-Octrees (SVO) are a ray-tracable hierarchical structure~\cite{SparseVoxelOctree}. Each node represents a voxel with potentially 8 child voxels. During rendering, we traverse from coarser to finer voxels until we either encounter a leaf node, or a voxel is small enough to stop the traversal and display it. Since cube-shaped voxels are not an ideal representation for curved surfaces, the authors also propose contour-based voxel representations that align with a surface's orientation. Crassin and Green demonstrate voxelization algorithms for the GPU that can also generate such SVOs by first expanding the hirearchy in a top-down fashion, then populating the leaf nodes with voxels, and eventually inner nodes via mip mapping~\cite{CG12}. Kämpe et al. later proposed directed acyclic graphs (DAG) as a more efficent encoding for SVOs that reduses the required memory by 1-2 orders of magnitude~\cite{HighresSVD}.  Beart et al.~\cite{OutOfCoreSVO} and Pätzold and Kolb~\cite{GridFreeSVOGen} introduced efficient out-of-core SVO construction procedures (e.g. input: San Miguel, 10M triangles, 4096 voxel grid, \textbf{SVO}: 153s, \textbf{Baert}: 26s, \textbf{Pätzold}: 12s). 

GigaVoxels~\cite{GigaVoxels} utilize an $N^3$ tree with $M^3$ bricks -- Each node splits into $N^3$ child nodes (an octree in case of $N = 2$), and nodes comprise bricks of $M^3$ voxels, with $M$ around 32. At runtime, nodes are traversed to find the voxel blocks with appropriate resolution which are then ray-marched. In addition, GigaVoxels can also filter colors from multiple brick levels to reduce aliasing artifacts via voxel-based mip mapping. Some similarities to the structure we generate are: we use an octree, hence $N=2$. We use chunks/bricks of $M^3=128^3$. The main difference is that we intent to rasterize sparse surfacic data, so instead of volumetric bricks, we generate vertex buffers with voxel coordinates and colors. 

\textbf{Rasterization-based LOD}: FarVoxels~\cite{FarVoxels} uses a BSP-tree of view-dependent voxels in inner nodes and triangles in leaf nodes. View-dependent meaning that shading parameters for various view directions are computed during construction of the LOD structure and stored in each voxel. Chajdas et al. also use a hybrid voxel-triangle LOD structure~\cite{Chajdas2014Scalable} but using an octree where each node represents a chunk of $256^3$ voxels, and leaf nodes store original triangle data. During rendering, the chunks with the lowest yet sufficient resolution are rendered. Once the viewer zooms in, however, even the leaf node's voxel resolution may not be sufficient. In that case, the original triangle data contained in that leaf node will be rendered instead. The structure that we generate in this paper closely matches the structure of Chajdas et al., with two minor differences: First, instead of $256^3$ cells in each node/chunk, we decided to use $128^3$ for each node as it reduces the amount of memory required during construction, as well as the amount of voxels in each octree node. Second, since we use point clouds, leaf nodes contain the original point instead of triangle data, and they do not contain any voxels. 

\subsection{Counting Sort}

Counting sort is a linear-time $O(n)$ integer sorting algorithm that is capable of efficiently sorting large amounts of integers with just two iterations over the entire input data~\cite{Knuth1998, Cormen2001}. In contrast, comparison-sort algorithms such as quicksort or merge sort have a complexity of $O(n \, \log n)$, and therefore the number of times each element of the input is accessed is not constant and increases logarithmically with the total length of the input array~\cite{Knuth1998}. The improved performance of counting sort comes with two limitations: First, it is limited to sorting integer keys. Second, the keys have a limited range because the algorithm maintains a counter for each potential key. It can therefore be thought of as sorting integers into \emph{buckets}. In this paper, we use counting sort to partition points into leaf nodes (the buckets) using the target node's voxel grid coordinate as the integer key.

\section{Data Structure}

The generated data structure is a multi-resolution octree where leaf nodes contain the original points, and inner nodes are made up of voxels at coarse, level-dependent resolutions, as shown in Figure~\ref{fig:teaser}. Unlike fine-grained acceleration structures that have ray-tracing in mind (e.g. SVOs), our structure follows the rasterization-friendly layered-point-cloud scheme~\cite{GOBBETTI2004}. As such, each node comprises a larger batch of points or voxels, which can be efficiently rasterized in a single draw call. To be precise, leaf nodes store up to $T = 50$k points, and inner nodes contain surface voxels that match an inscribed $128^3$ grid. Since we target surfacic 3D models, the number of occupied voxels in such a grid is usually closer to $128^2$, so instead of storing the entire, sparsely populated grid, we store a vertex buffer comprising coordinates and colors of occupied voxels. An important difference between our structure and regular LPCs is that LPCs distribute all points to all nodes of the entire tree structure, whereas our structure partitions all points into the leaf nodes. LPCs do not create additional or duplicate data -- the original point cloud is recovered by merging points in all nodes of all levels. Our structure, on the other hand, creates new voxel data in inner nodes that is meant to faithfully represent the original model at specific resolutions. In that regard it is similar to the structure of Wand et al.~\cite{Wand2008}, who store original point data in leaf nodes and surfels in inner nodes. It is also similar to the structure of Chajdas et al.~\cite{Chajdas2014Scalable}, who store triangles in leaf nodes and chunks of $256^3$ voxels in inner nodes. 

\section{LOD Construction}

Our CUDA-based LOD construction method creates the aforementioned LOD structure in two main steps: First, we partition all points of the input data set into octree leaf nodes with at most $T = 50k$ points. Second, we populate inner nodes with lower-resolution voxel models from bottom-up. By splitting into leaf-nodes with $T$ points, we generate spatially constrained work packages that can be processed by numerous workgroups in parallel. 

\begin{figure}
    \centering
    \includegraphics[width=\columnwidth]{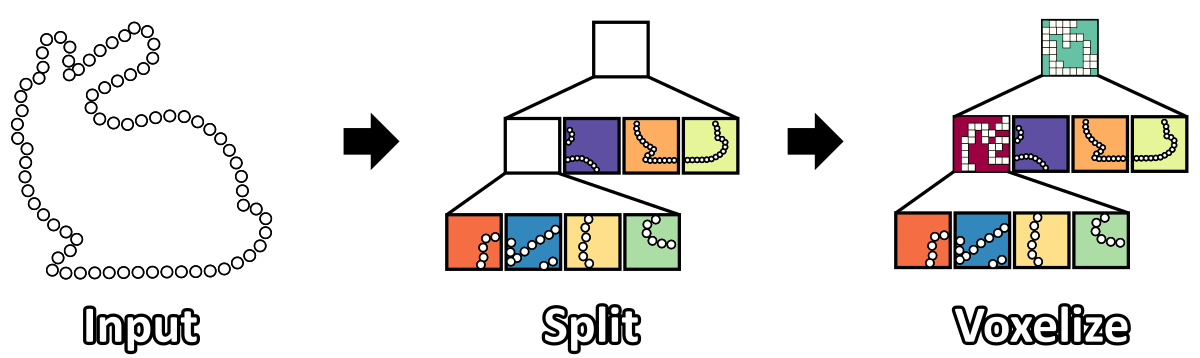}
    \caption{LOD Construction Overview: First, points are split into leaf nodes of an octree with hierarchical counting sort. Then, inner nodes are populated bottom-up with voxelized representations of their children.}
    \label{fig:method_overview}
\end{figure}

\begin{figure}
    \centering
    \begin{subfigure}[t]{\columnwidth}
        \includegraphics[width=\textwidth]{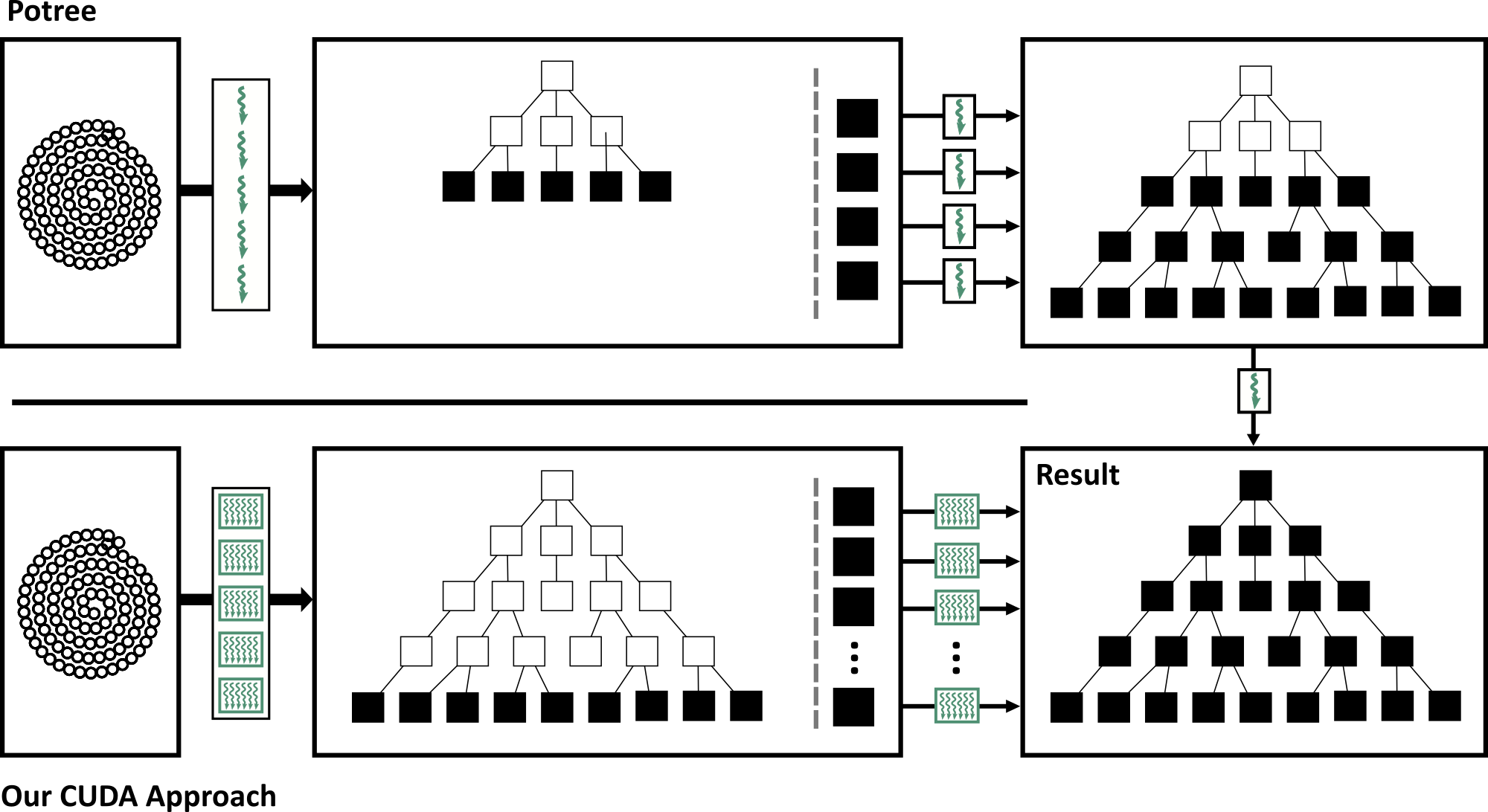}
    \end{subfigure}
    \caption{Parallelism in Potree vs our CUDA-based approach. Potree uses all CPU threads to split the octree into chunks of 10M points. Chunks are then further split and converted into an octree, using one thread per chunk. A single thread then merges all chunks and populates the remaining coarsest LODs. Our CUDA-based approach directly splits into leaf nodes with about 50k points utilizing all threads in all workgroups, and it then populates inner nodes utilizing one workgroup per node.}
    \label{fig:parallelism}
\end{figure}

\subsection{Split Input into Leaf Nodes}

\begin{figure}
    \centering
    \begin{subfigure}[t]{\columnwidth}
        \includegraphics[width=\textwidth]{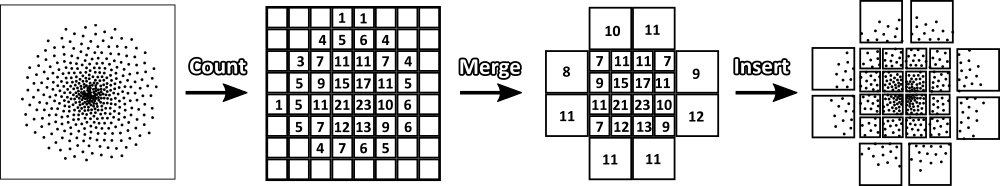}
    \end{subfigure}
    \caption{We directly create an up to 8 levels deep octree using a hierarchical counting sort approach. Counting and merging gives us the full octree hierarchy with knowledge about the number of points in leaf nodes. We then allocate the right amount of memory and insert points into the respective leaves.}
    \label{fig:method_split}
\end{figure}

To partition the input data into octree leaf nodes with at most $T$ points, we adapt the hierarchical counting sort approach of Schütz et al.~\cite{SCHUETZ-2020-MPC}. Hierarchical counting sort allows us to partition a point cloud into an octree with $depth$ levels by iterating over all points just twice -- once to count and a second time to move the points to the sorted location in memory. For this CUDA approach, we suggest an initial octree depth of 8 (recursively extended, if needed). 

\subsubsection{Counting}

At octree level 8, we have $(2^{level})^3 = (2^8)^3 = 256^3$ potential leaf nodes, so we create a corresponding counting grid with $256^3$ cells. We then iterate through all points, project them into the counting grid and atomically increase the counter of the corresponding cell. 

\subsubsection{Merging}

Although we want our leaf nodes to contain at most $T$ points, we also do not want them to contain too few points, so in this step, we merge all groups of 2x2x2 cells that contain less than $T$ points, as shown in Figure~\ref{fig:method_split}. For this purpose, we create a pyramid of counting grids with a size of $(2^{level})^3$ for each octree level. Whenever a 2x2x2 group of cells contains less than $T$ points, we store the sum of points in the next coarser counting grid, and set all involved cells of the current level to 0. In addition to that, we also flag the cell at the coarser level as \emph{unmergeable} if we were not able to merge it because the sum of points exceeded the threshold. Furthermore, if a 2x2x2 group of cells contains at least one cell that was previously marked as \emph{unmergable}, then the whole group becomes unmergeable and is also flagged as such in the coarser level. 

\subsubsection{Creating Target Pointers}

Each non-zero value in the counter pyramid represents an octree node. Positive integer values represent a leaf node and the amount of points they contain, while cells that were flagged as \emph{unmergeable} represent inner nodes that will be populated with a currently unknown amount of voxels at a later time. We now create a list of all leaf and inner nodes, and in case of leaf-nodes, we allocate sufficient memory for the points. Additionally, we map the counter pyramid into a pointer pyramid, where each cell points to the respective leaf node (counter > 0), inner node (counter = \emph{unmergeable}), or null if the cell is empty (counter = 0). The process of creating the target pointers corresponds to creating the prefix sums in regular counting sort. In both cases, the results are used to compute the sorted position of points in the final step. 

\subsubsection{Insert}

The last step of the node-wise sorting procedure iterates over all points a second time, projects them again to cells of a $256^3$ grid, but this time each point reads the node pointer from the recently created pointer pyramid. If we encounter a non-null pointer, we found the leaf node that this point belongs to, and add it. If we encounter a null pointer, then the cells at this octree level were merged into a lower level, i.e., we iterate upwards in our pointer-pyramid until we eventually encounter the leaf-node pointer. 

\subsubsection{Recursive Partitioning (If Necessary)}

After these steps, all points are sorted into an octree with a depth of at most 8 levels, or fewer if small nodes were merged. However, 8 levels are not always sufficient, especially with massive amounts of points, or irregularly distributed points with dense clusters in some regions (e.g. teapot-in-a-stadium scenario). This issue is easily addressed by repeating the sort procedure for all leaf nodes that are still too large. Additional iterations can also adjust the sort-depth and counting-grid size to fit the computational effort.

In our case, we implemented one level of recursion (sufficient for all test data sets) in-between the initial counting and the merging step. After the initial counting procedure, some cell counters may exceed the given threshold $T$. For each such cell, we generate an additional counting grid pyramid with an octree depth of 4, comprising $(2^4)^3 = 16^3$ cells at the highest level and extending the octree from the initial up to 8 levels to up to 12 levels. For all cells that exceeded the threshold, we replace the counter in the main counting grid with a pointer to the newly created extended counting grid. We then iterate through all points again, project them to the original grid, and if a point hits a cell with a pointer, we follow the pointer and update the counters in the extended counting grid. 

The merging, pointer creation and insertion steps are modified correspondingly. The merging step additionally merges the counters in the extended counting grids, and the pointer creation step also creates octree nodes and pointers for the extended grids. The insertion step still projects points to the main grid with depth 8, size $256^3$, but additionally checks whether the targeted cell contains a regular node pointer or a pointer to an extended grid. If it's a pointer to an node, we proceed as usual and add the point to that node. If it's a pointer to an extended grid, we traverse into that grid and look there for the corresponding octree node pointer.

\subsubsection{Implementation Details}

Implementation-wise, the split procedure was realized in a single CUDA kernel that heavily uses cooperative groups. The amount of available workgroups and threads remains static for all passes of the kernel (clearing buffers, counting, merging, recursion, ...) and the varying amount of work items (points, count grid cells, unknown amount of extended grids, ...) is dynamically distributed to the fixed amount of threads and workgroups. Global syncing of all threads with the cooperative group API (\emph{grid.sync()}) ensures that one pass finishes before another one uses the results. Although processing the extended counting grids appears to be an opportunity for workgroup-based processing of each grid, we found it to be more efficient to pack all extended grids into a single buffer and treat them like a single, large array of cells. This ensures that all GPU threads are utilized evenly, even if the number of extended grids is smaller than the number of streaming multiprocessors. Individual threads get the index of the cell they should process, and they can identify the corresponding extended grid via $extendedGridIndex = \frac{cellIndex}{numCellsPerGrid}$.

\subsection{Voxel Sampling}

\begin{figure}
    \centering
    \begin{subfigure}[t]{0.28\columnwidth}
        \includegraphics[width=\textwidth]{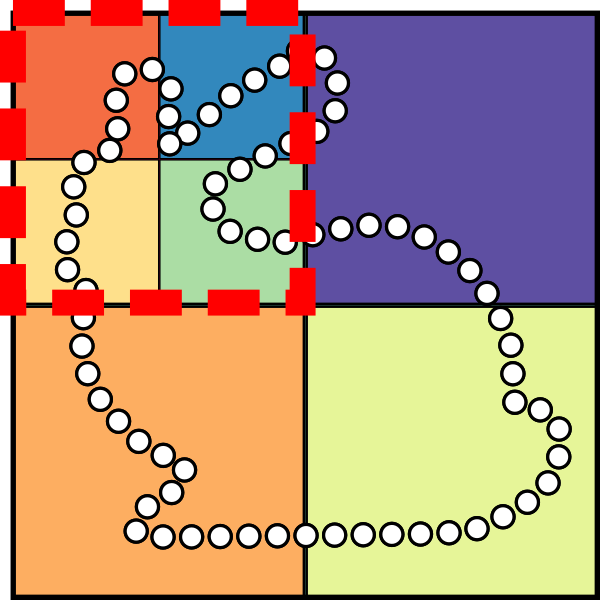}
        \caption{}
    \end{subfigure}
    \hfill
    \begin{subfigure}[t]{0.28\columnwidth}
        \includegraphics[width=\textwidth]{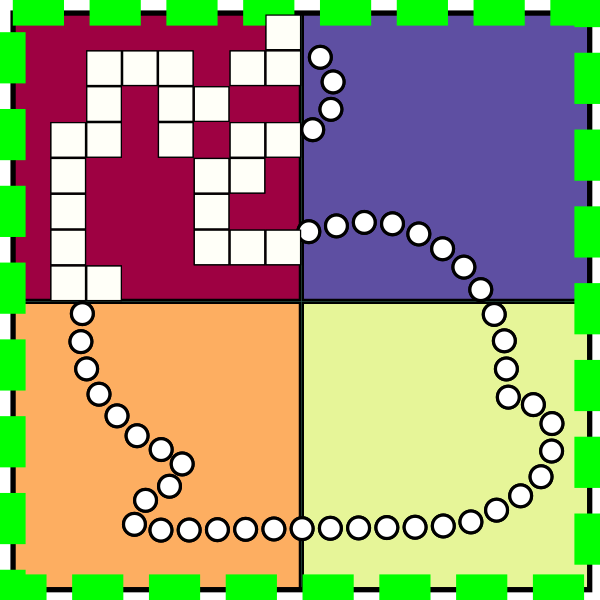}
        \caption{}
    \end{subfigure}
    \hfill
    \begin{subfigure}[t]{0.28\columnwidth}
        \includegraphics[width=\textwidth]{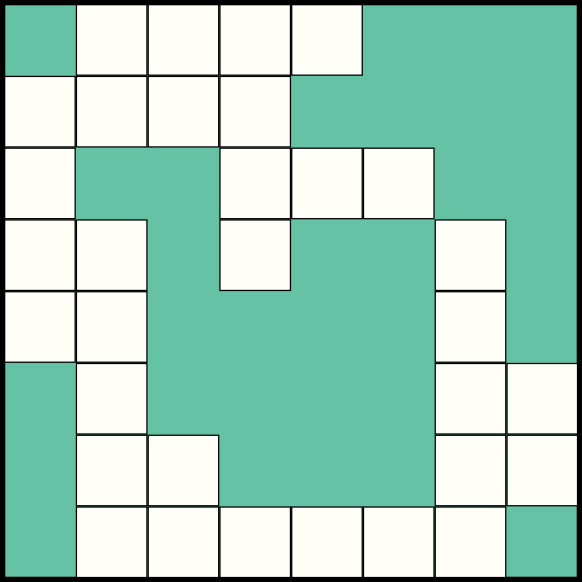}
        \caption{}
    \end{subfigure}
    \begin{subfigure}[t]{0.28\columnwidth}
        \includegraphics[width=\textwidth]{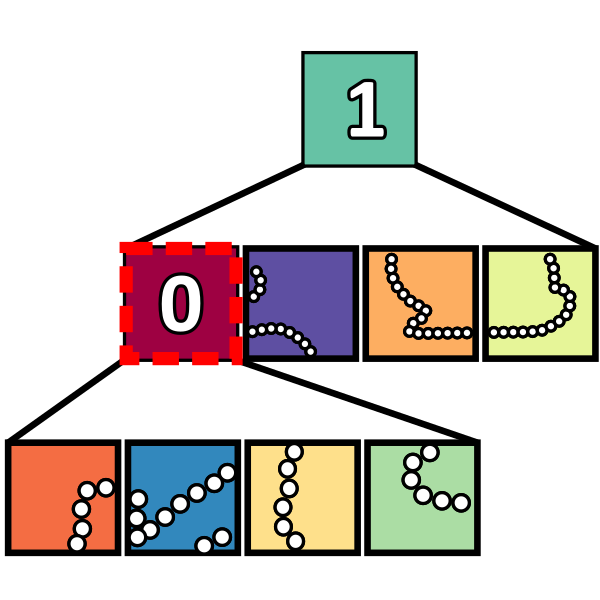}
        \caption{}
    \end{subfigure}
    \hfill
    \begin{subfigure}[t]{0.28\columnwidth}
        \includegraphics[width=\textwidth]{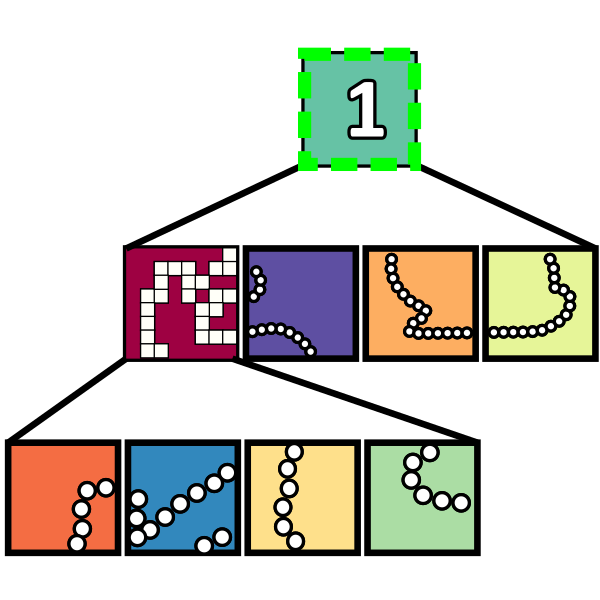}
        \caption{}
    \end{subfigure}
    \hfill
    \begin{subfigure}[t]{0.28\columnwidth}
        \includegraphics[width=\textwidth]{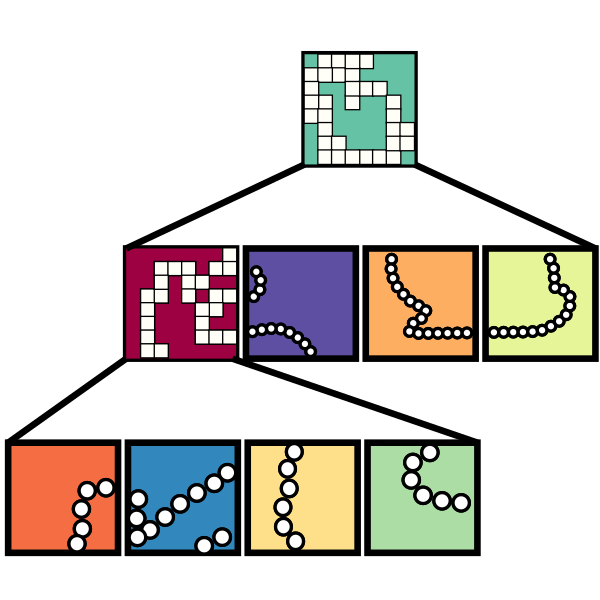}
        \caption{}
    \end{subfigure}
    \caption{Creating lower LODs: (a+d) The split pass partitioned all points into leaf nodes. Node $0$ (dashed red) will be filled with a voxelized representation of all its children. (b+e) Node $1$ (dashed green) is populated next with a coarser voxelized representation of voxels and points from its child nodes. (c) The voxelized root node. (f) Tree hierarchy showing leaf nodes with points and inner nodes with voxels.}
    \label{fig:voxelization}
\end{figure}

\begin{figure}
    \begin{subfigure}[t]{0.2\columnwidth}
        \includegraphics[width=\textwidth]{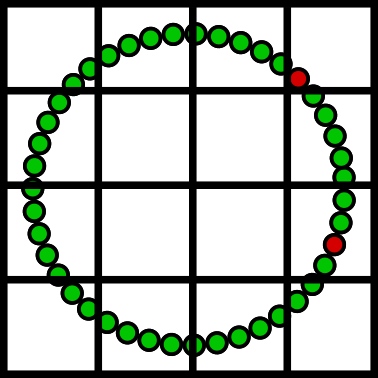}
        \caption{}
    \end{subfigure}
    \hfill
    \begin{subfigure}[t]{0.2\columnwidth}
        \includegraphics[width=\textwidth]{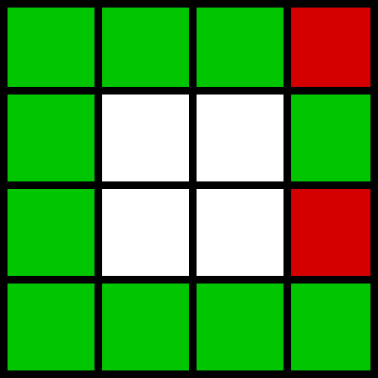}
        \caption{}
        \label{fig:sampling_pick}
    \end{subfigure}
    \hfill
    \begin{subfigure}[t]{0.2\columnwidth}
        \includegraphics[width=\textwidth]{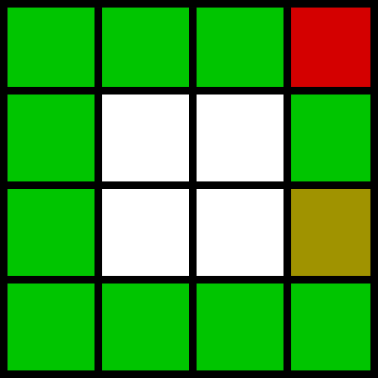}
        \caption{}
        \label{fig:sampling_average}
    \end{subfigure}
    \hfill
    \begin{subfigure}[t]{0.2\columnwidth}
        \includegraphics[width=\textwidth]{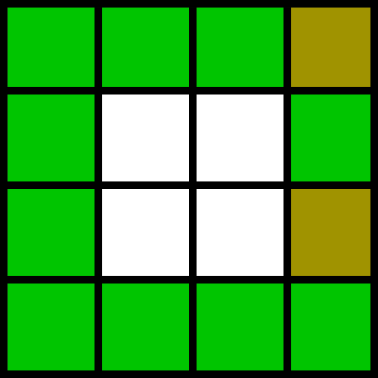}
        \caption{}
        \label{fig:sampling_weighted}
    \end{subfigure}
    \caption{(a) Samples and sampling grid. (b) Selecting a single sample per cell leads to a disproportional representation of outliers. (c) Averaging within a cell can improve color quality through proportional representation. Still susceptible to outliers from partial intersections. (d) Weighted averaging of samples within a distance, including those in adjacent voxels, further improves quality.}
    \label{fig:sampling}
\end{figure}

After partitioning the input point cloud to the leaf nodes of an octree, we proceed to populate lower LODs from bottom up with voxelized representations of higher LODs. We start by creating a list of the bottom-most empty inner nodes with exclusively non-empty child nodes, and then populate each node in this list with a coarser, voxelized representation of its child nodes, using one workgroup per node. This process of finding the bottom-most empty nodes and populating them is repeated until we eventually process the root node and thereby finish the LOD construction. 

%We recursively iterate through the bottom-most inner nodes, specifically nodes whose children are all non-empty, i.e., already filled with either points or voxels. For each such inner node, we create a lower-resolution voxel representation based on the points and voxels of its children. 

Like splitting, the entire voxel-sampling kernel is implemented inside a single CUDA kernel. Unlike splitting, however, it makes heavy use of per-workgroup processing of work packages. Each workgroup fetches and processes one node at a time, until all nodes up to the root are populated with voxels. Reasons for matching nodes with workgroups are: We only need to allocate one sampling grid per workgroup; nodes frequently have less points or voxels than we have available GPU threads, so utilizing more threads is wasted effort; and we can take advantage of shared memory (L1 cache). Furthermore, each workgroup requires a considerable amount of global memory for the voxel sampling grid, so in order to minimize the amount of required memory while still utilizing all streaming multiprocessors, we launch exactly one workgroup per SM.

%Furthermore, we spawn exactly one workgroup per streaming multiprocessor in order to keep all SMs busy while simultaneously minimizing the amount of voxel sampling grids we need to allocate for the workgroups. 

There are several possible sampling strategies that can be used to create a lower-resolution voxel representation out of the points and voxels in child nodes. They all have in common that after fetching a node to be voxelized, they iterate through the points and/or voxels of the 8 child nodes, project them to a sampling grid with a dimension of $128^3$ cells, and finally extract the voxel samples from the grid and store them inside the node. They differ in the details of the sampling grid and the extraction process. We investigated the following sampling strategies, which provide trade-offs in construction performance and quality:

\subsubsection{Sampling Strategy: First-Come-First-Serve}

This fastest approach accepts the first sample (point or voxel) in each sample grid cell. The sample grid of the node we currently populate has a size of $128^3$, but this approach processes each of its 8 child nodes sequentially, thus we only need to use and reuse a $64^3$ grid for each child. The grid is implemented as a bitmask in shared memory, requiring $64^3 = 262$k bits, equal to 32 KB, which nicely fits into the limit of 48 KB static shared memory on CUDA devices. We also allocate a temporary list of accepted samples with a static capacity of $C$ (e.g., 200k -- more than we ever expect a node to accept) and a counter that tracks how many samples we accepted. This temporary, large list is used as an intermediate storage for a node's voxels, because we do not know the exact amount of required memory for each node in advance. The list is reset after each processed node by setting the counter to 0.

Each workgroup then fetches a node from a list of unprocessed nodes, iterates through that node's 8 children, first clearing the sample grid at the start of iteration, then projecting the child's samples (points or voxels) to the grid and atomically setting the corresponding cell's bit to 1 via \emph{atomicOr}. The return value of \emph{atomicOr} reports the previous value -- if the cell's bit was previously 0 and is now 1, then the sample is the first processed sample to occupy that cell. In that case, we accept it and add it to a temporary list of accepted samples. If it is not the first (the corresponding bit was already 1), we ignore it. After iterating through all child nodes and their samples, we create an array of voxels as large as the amount of accepted samples to the currently processed node. The accepted samples are then voxelized by quantizing their coordinates to the $128^3$ sampling grid, and storing them in the newly allocated array.

%After iterating through all child nodes and their samples, we allocate as much memory as we need for the amount of accepted samples, and copy the samples from the temporary buffer. Finally, we update the node's sample counter and its pointer to the accepted voxel samples. 

\subsubsection{Sampling Strategy: Random}

For each cell in our voxel sampling grid, we accept one random sample (point or voxel) from the child nodes. Choosing the random sample efficiently is done by computing a random number for each sample, encoding random number and sample index into a 32 bit integer, and then using atomicMax on the sample grid cell to retain the sample with the largest random value, as shown in Listing~\ref{lst:randomselect}.

\begin{lstlisting}[language=Java,label={lst:randomselect},caption={CUDA-pseudocode that selects one random sample and stores its ID inside a voxel cell.},captionpos=b]
uint32_t encoded = 
    (randomNumber & 0xfff00000) |
    (sampleIndex  & 0x000fffff);
atomicMax(&voxelGrid[voxelIndex], encoded);
\end{lstlisting}

Since we now require 32 times as many bits per cell (for the 32 bit integer instead of 1 bit occupancy), the sampling grid does not fit in shared memory anymore. Instead, we allocate one $128^3$ voxel sampling grid per workgroup in global memory, which is reused for every node a workgroup processes. The amount of global memory required for each workgroup is $sizeof(uint32\_t) * 128^3 = 8$MB. To minimize the total amount of global memory reserved by all workgroups, we only launch one workgroup per streaming multiprocessor for a total of $SMs * sizeof(uint32\_t) * 128^3$ bytes, which is equal to 687MB on an RTX 3090 with 82 SMs. Similar to the first-come strategy, we track whenever an empty cell is filled with a sample, but since the accepted sample in a cell may randomly change, we create a list of occupied voxels instead of a list of accepted samples. After all points and voxels in the child nodes were projected to the sampling grid, we allocate a list of voxels with a size equal to the number of occupied voxel cells for the current node; then iterate through the list of occupied voxels; retrieve the indices of the accepted samples from the voxel grid; and finally store the respective sample in the node's list of voxels. Note that each sampling grid ($128^3$ cells) only needs to be fully cleared once at the beginning, but for reuse we only need to clear the voxels that were occupied (closer to $128^2$ cells).

\subsubsection{Sampling Strategy: Cell-wise Average}

A single-sample color value (random, first-come, etc.) is a poor representation of all the higher-LOD samples that a voxel should represent, as shown in Figure~\ref{fig:sampling_pick}. This strategy generates a more representative color by computing the average value from all samples that are projected to a voxel cell, as shown in Figure~\ref{fig:sampling_average}. For this strategy, each workgroup allocates a $128^3$ sampling grid in global memory. Each cell stores the sum of red, green and blue values, as well as a counter of samples that contributed, thus each workgoup requires $4 * sizeof(uint32\_t) * 128^3$ bytes, i.e., 33MB. Since we spawn one workgroup per streaming multiprocessor, the kernel requires a total of $SMs * 33$ MB, about 2.7 GB on an RTX 3090 with 82 SMs. Like the random strategy, this strategy also maintains a list of occupied voxel cells and corresponding counter.

Each workgroup then fetches the next unprocessed node, iterates through all samples in that node's children and projects them to the sample grid. The RGB values of the sample are atomically added to the RGB values of the corresponding sample grid cell, and the counter of the sample grid cell is incremented by one. The first sample that incremented a cell's counter also adds that cell's voxel index to the list of occupied voxels. Afterwards, we allocate sufficient memory for an array of occupied voxels, then iterate through the list of occupied voxels, retrieve that voxel's RGB values and counter and compute the arithmetic average. The voxel coordinate and average color are then stored inside that node's list of voxels. As with the random sampling strategy, we only need to clear the occupied voxels in the sampling grid.

\subsubsection{Sampling Strategy: Neighborhood Weighted Average}

Computing the arithmetic mean of colors within a voxel is fast and improves quality, but it still suffers from aliasing issues, especially when a voxel captures a single sample that strongly differs from adjacent samples, as shown in Figure~\ref{fig:sampling_average}. This strategy further improves quality by computing the weighted average of all samples within a certain distance to the center of a voxel, as shown in Figure~\ref{fig:sampling_weighted}. The closer to the voxel center, the higher the contribution. In this paper, we use a simple linear weight function that cuts off at a distance of 1 (=width of a cell):

\begin{lstlisting}[language=Java]
distance = length(samplePosition - voxelCenter)
weight   = clamp(1 - distance, 0, 1)
\end{lstlisting}

Coordinates are relative to the voxel sampling grid, i.e., $[0, 128)$. The given weight function evaluates to 1 if the sample position is equal to the target voxel position, or smaller than 1 otherwise. If a sample is farther than one cell's width away from the voxel center, the weight is cut to zero. Therefore, each sample may affect at most 2x2x2 surrounding voxels and we need to project each sample to 8 surrounding voxels, compute the corresponding weights, and add the weighted color values as well as the weight itself to the voxel grid using atomicAdd. 

Although each sample may affect color values of up to 8 voxels, only one of these voxels actually contains that sample. That respective voxel is flagged as \emph{occupied}. In the next step, we iterate over all $128^3$ cells of the voxel sampling grid and extract those cells that are marked as occupied. Non-occupied cells are deliberately ignored because they would dilate the model. During extraction, we create a list of voxels, where a voxel's coordinate is the coordinate of an occupied cell, and a voxel's color is that cell's sum of colors, divided by its sum of weights.

The result of the LOD construction procedure is a voxelized LPC~\cite{GOBBETTI2004} (or voxel-point-based version of Chajdas et al. \cite{Chajdas2014Scalable}) where leaf nodes comprise up to $T$ points, and inner nodes comprise lists of voxels that were generated by downsampling on a $128^3$ sampling grid.

\section{Rendering}

The proposed voxelized LPC structure can be rendered in largely the same manner as other layered point-cloud approaches~\cite{GOBBETTI2004, scheiblauer2011, SCHUETZ-2020-MPC}. Each voxel is treated as a point, and we aim for a level of detail where voxels are roughly pixel-sized. We then traverse the octree, invoke draw calls for all nodes that are within the view-frustum and whose projected bounding box is larger than about 100 pixels (matching the voxel-grid size of $128^3$, and the goal to render pixel-sized voxels). 

The main difference to other LPC approaches stems from the fact that LPCs are traditionally additive LOD approaches, meaning that higher levels do not replace lower levels -- they are added and rendered together. Our voxelized LPC, on the other hand, is a replacing LOD structure where higher levels of detail completely replace lower levels, which has the advantage that lower-LOD voxels better represent the color values at their given level of detail and catch radius. As for rendering, we invoke one draw call for each visible node and pass the respective node as a uniform. The vertex shader then checks whether the currently processed vertex/voxel is located in an octant with a visible child node. If it is, the voxel is discared because the child node will replace it with finer voxels. If there is no visible child node, the voxel is drawn.

\section{Evaluation}
\label{sec:evaluation}

\begin{figure*}
	\centering
	\begin{subfigure}[t]{0.24\textwidth}
		\includegraphics[width=\textwidth]{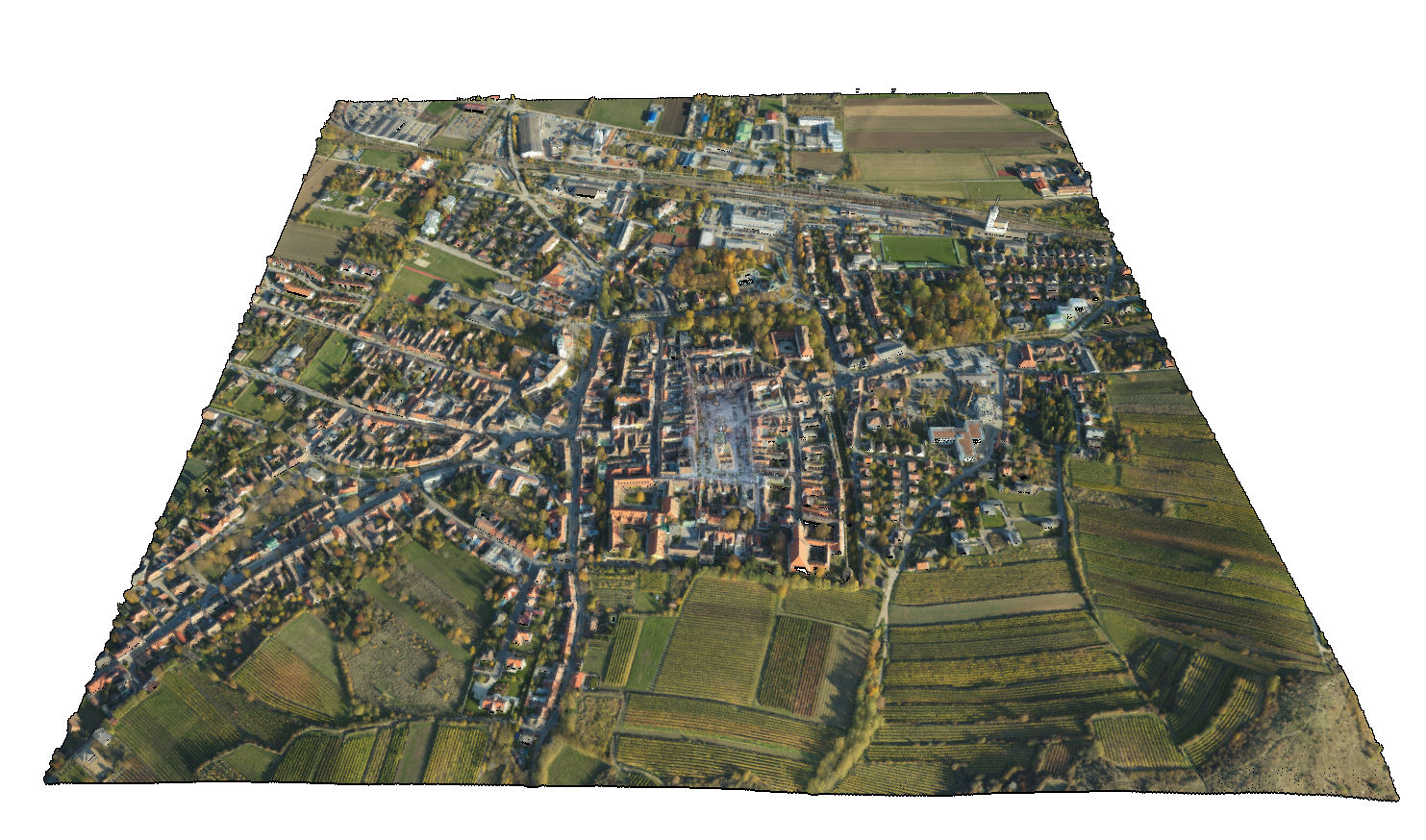}
		\caption{Retz}
	\end{subfigure}
	\hfill
	\begin{subfigure}[t]{0.24\textwidth}
		\includegraphics[width=\textwidth]{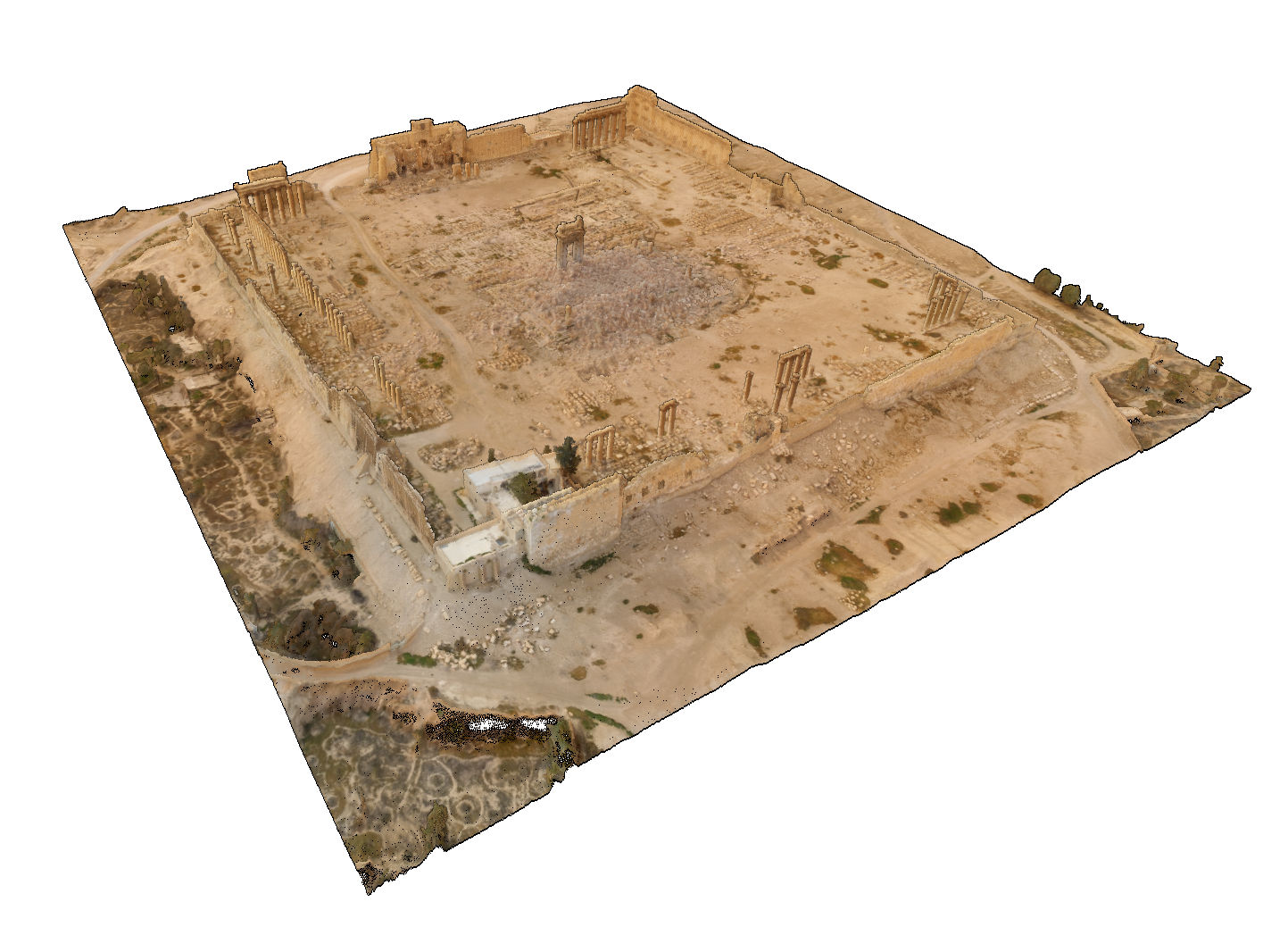}
		\caption{Palmyra}
	\end{subfigure}
	\hfill
	\begin{subfigure}[t]{0.24\textwidth}
		\includegraphics[width=\textwidth]{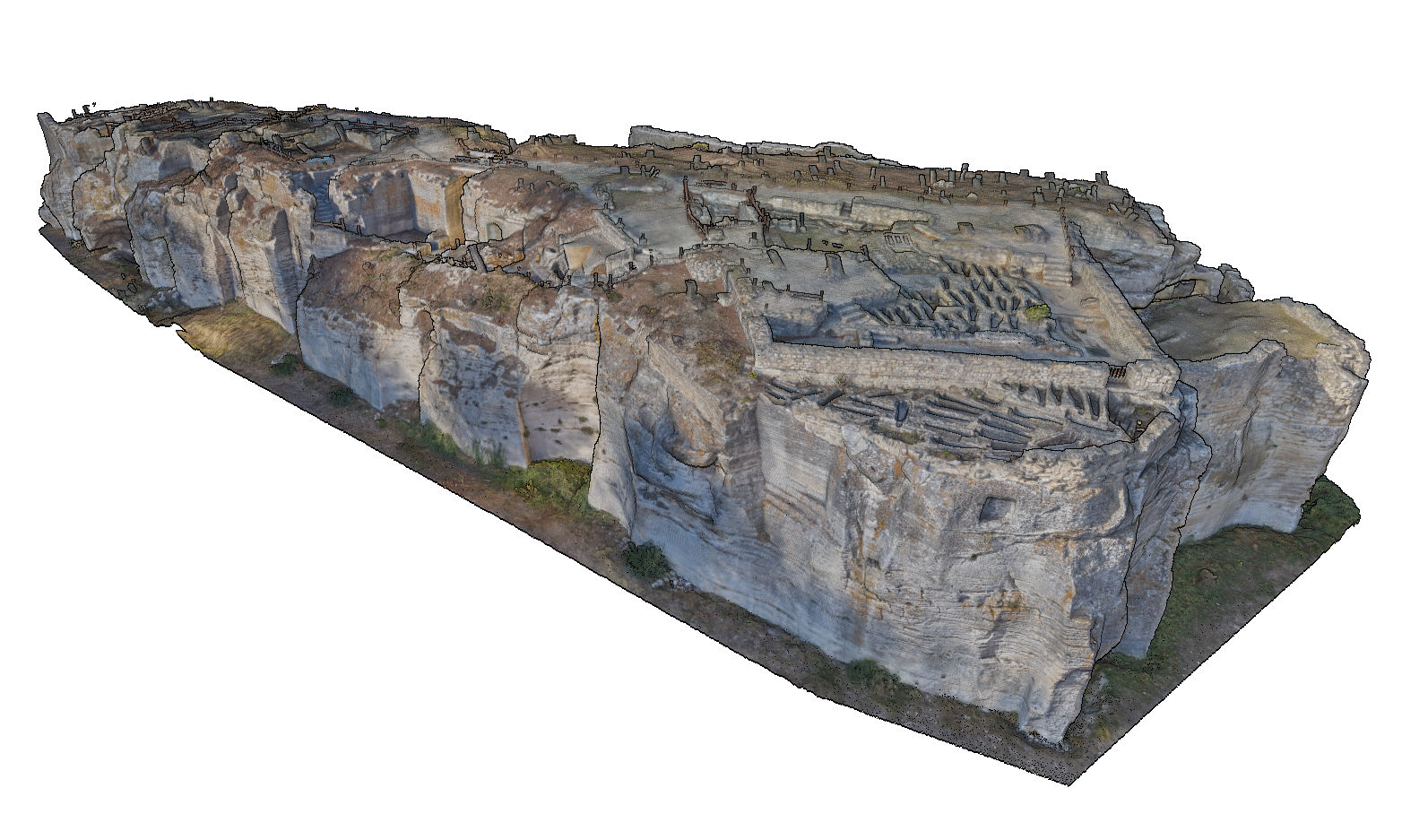}
		\caption{Saint Roman}
	\end{subfigure}
	\hfill
	\begin{subfigure}[t]{0.24\textwidth}
		\includegraphics[width=\textwidth]{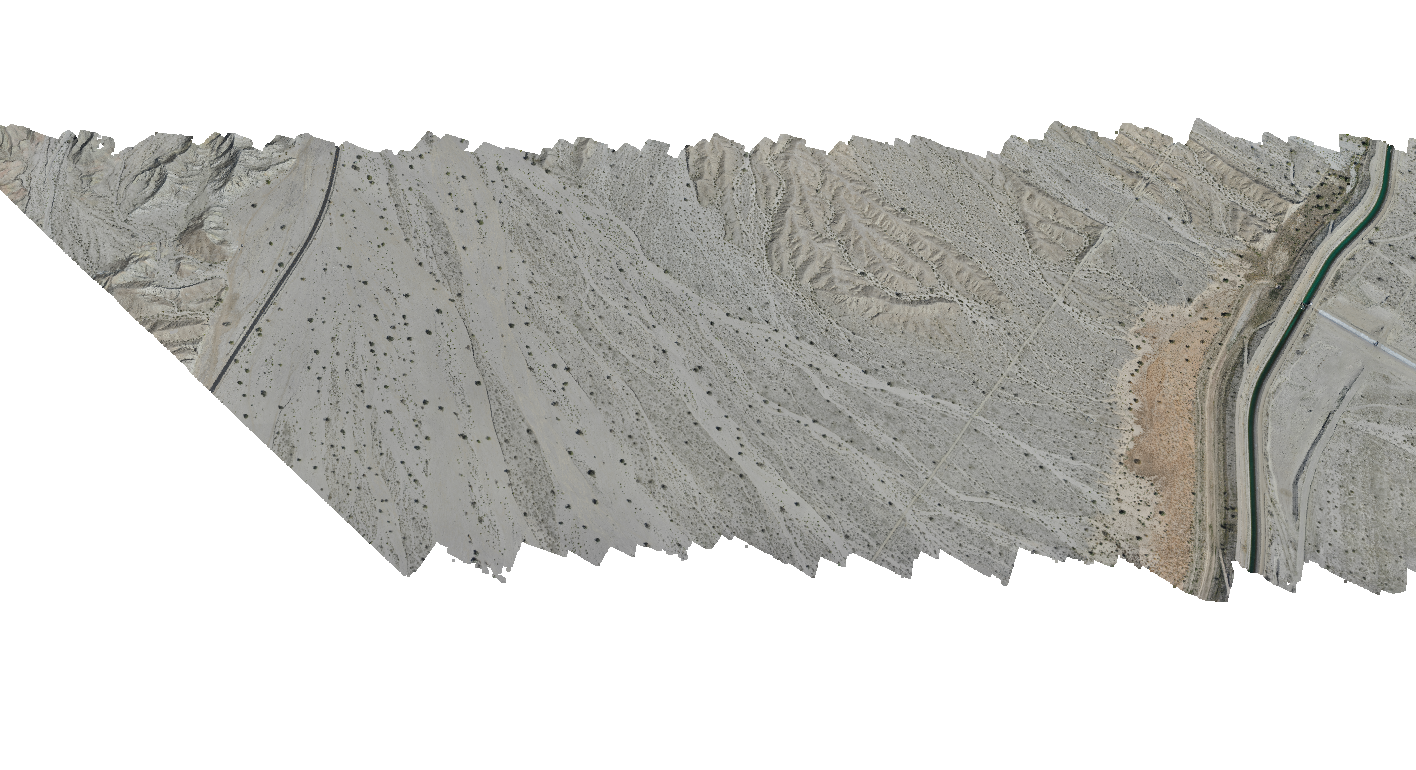}
		\caption{CA21 Bunds}
	\end{subfigure}
        %%%%%%
        % second row
        %%%%%%
	\begin{subfigure}[t]{0.24\textwidth}
		\includegraphics[width=\textwidth]{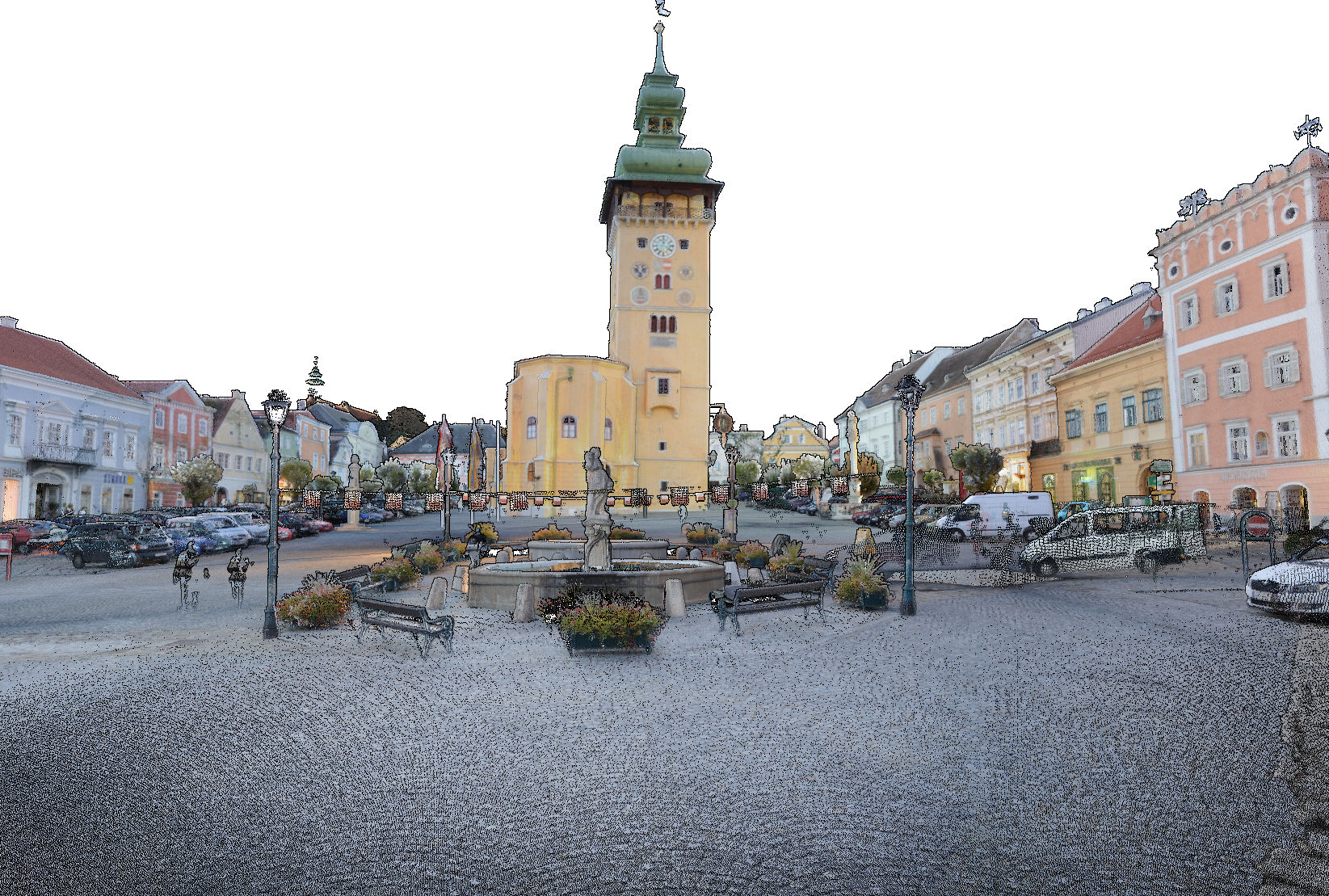}
		\caption{Retz}
	\end{subfigure}
        \begin{subfigure}[t]{0.24\textwidth}
		\includegraphics[width=\textwidth]{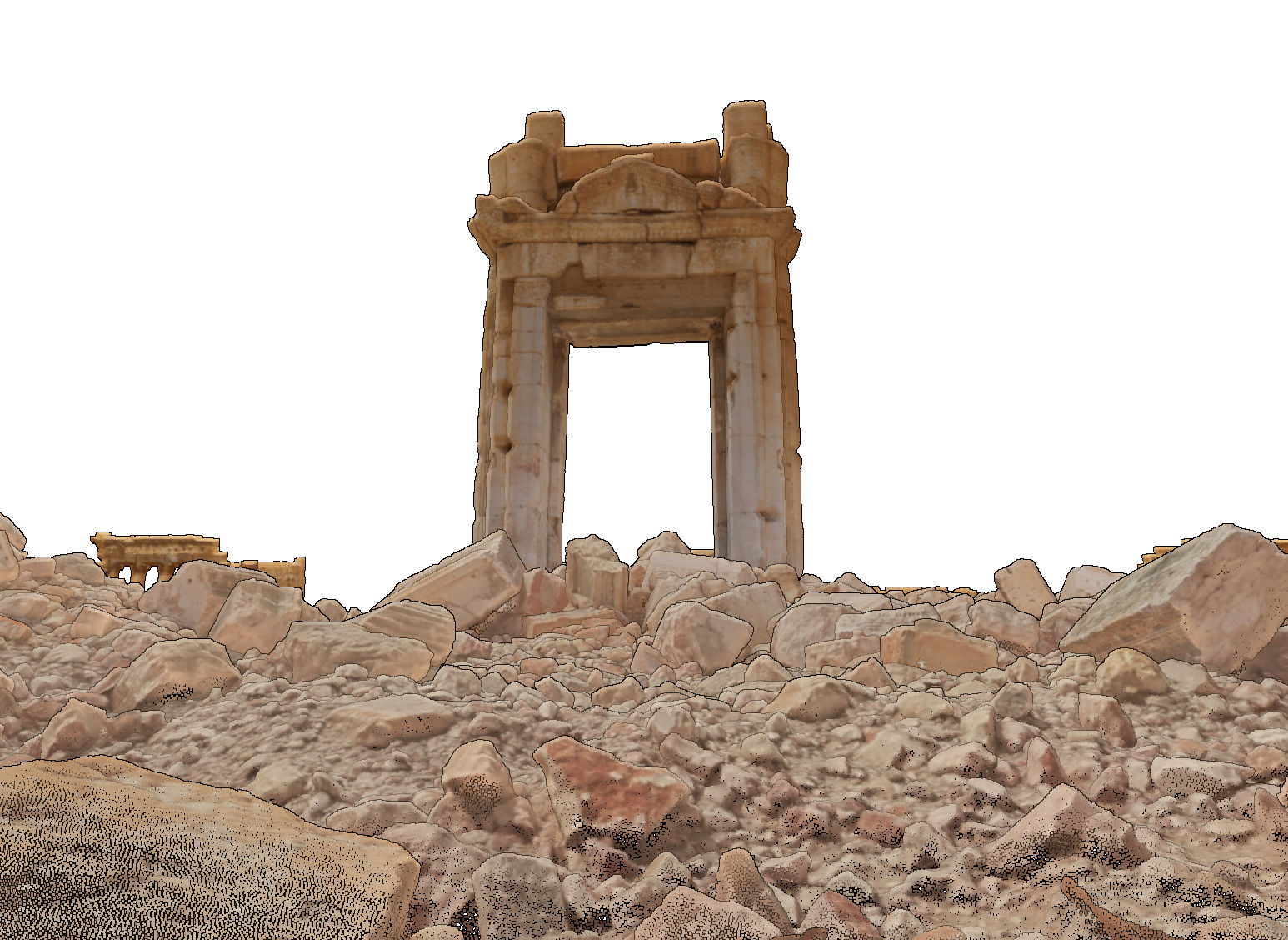}
		\caption{Palmyra}
	\end{subfigure}
	\hfill
	\begin{subfigure}[t]{0.24\textwidth}
		\includegraphics[width=\textwidth]{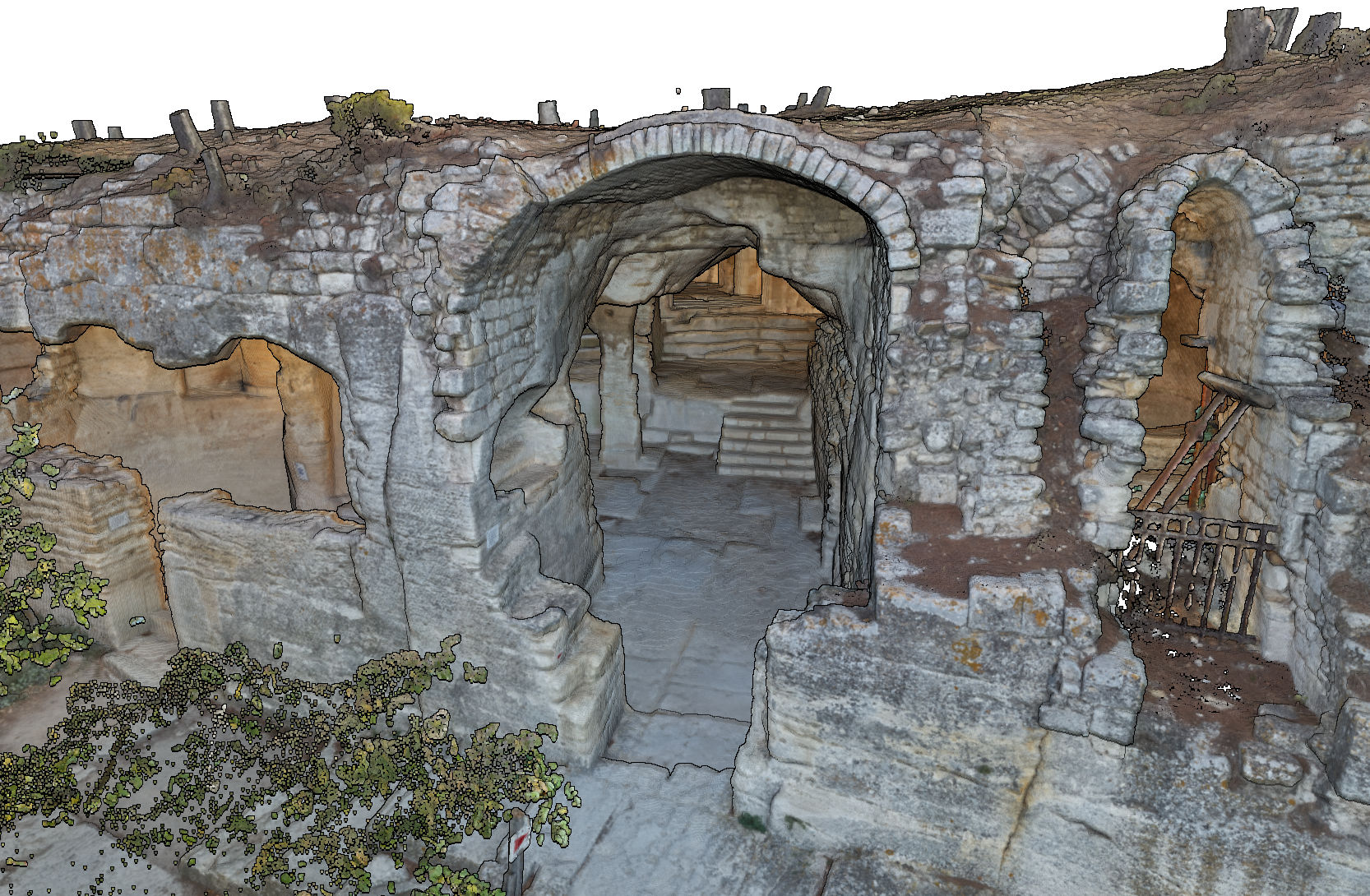}
		\caption{Saint Roman}
	\end{subfigure}
	\hfill
	\begin{subfigure}[t]{0.24\textwidth}
		\includegraphics[width=\textwidth]{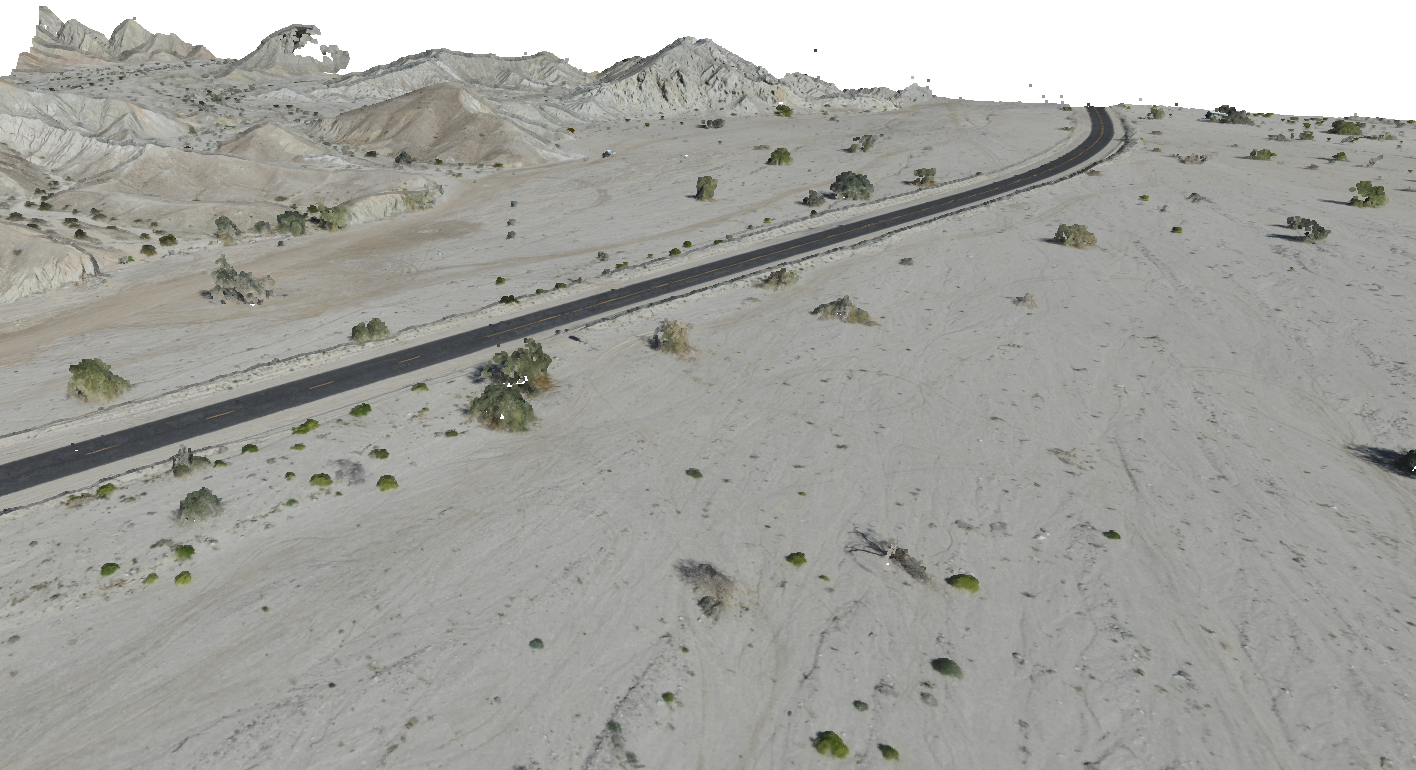}
		\caption{CA21 Bunds}
	\end{subfigure}
	\caption{Four data sets that challenge different aspects of the LOD construction process. Retz and Palmyra feature a teapot-in-a-stadium scenario (much higher point densities in regions of interest), Saint Roman comprises scans of outside as well as interior rooms, CA21 Bunds features a large amount of LIDAR data. }
	\label{fig:test_data_sets}
\end{figure*}

The proposed method was implemented in C++ and CUDA, and evaluated with the test data sets shown in Figure~\ref{fig:test_data_sets}. The data sets were chosen to demonstrate that our method works on various types of data, including mostly 2.5D-like aerial LIDAR scans, but also on more complex objects with higher depth complexity (hidden surfaces) and data sets where some regions have a much higher point density (teapot-in-a-stadium scenario). Kernel run times were measured with \emph{cuEventElapsedTime}. Only LOD construction times on the GPU were measured, I/O and host-device copies are not included. For each measure, we ran the construction process 5 times and report the median (to avoid highly volatile cold-start times). 

The evaluation was conducted on two GPUs for our CUDA-based approach, and one CPU to compare with the CPU-based state of the art (Potree~\cite{SCHUETZ-2020-MPC}).

\begin{itemize}
    \item NVIDIA RTX 3090 24GB %, AMD Ryzen 7 2700X (8 cores), 64GB RAM, running Windows 10.
    \item NVIDIA RTX A6000 48GB
    \item AMD Ryzen 7 2700X (8 cores), 64GB RAM, Windows 10. 
\end{itemize}

\subsection{LOD Construction Performance}

\begin{table*}
\begin{tabular*}{\textwidth}{@{\extracolsep{\fill}} |c|l|r|r|rrrr|rr|r|}
\hline 
 &                     &           &       & \multicolumn{4}{c|}{Ours}                                & \multicolumn{2}{c|}{Potree}     &            \\
 & Data Set            &  points   &  GPU  &    split     & voxelize &  total   &   MP/s                   &    total   &         MP/s    &  Speedup   \\
\hline                                                                                                                                               
 & Retz                &     145 M &  RTX 3090 &        20.0  &    30.1  &  50.1    &    2\,906           &       13\,130 &          11     &   $\times$ 262      \\
 & Palmyra             &     258 M &  RTX 3090 &        36.0  &    67.4  & 103.4    &    2\,504           &       21\,269 &          12     &   $\times$ 205      \\
 & Saint Roman         &     547 M &  RTX 3090 &        75.3  &   137.7  & 213.0    &    2\,569           &       43\,570 &          13     &   $\times$ 204      \\
 \rot{\rlap{~random}}
 & CA21\_Bunds         &     976 M &  RTX A6000   &       133.1  &   131.6  & 264.8    &    3\,685           &       85\,651 &          11     &   $\times$ 323      \\
\hline 
 & Retz                &     145 M &  RTX 3090 &        20.8  &    12.8  &  33.7    &    4\,319           &       13\,130 &          11     &   $\times$ 389      \\
 & Palmyra             &     258 M &  RTX 3090 &        37.1  &    19.0  &  56.1    &    4\,619           &       21\,269 &          12     &   $\times$ 379      \\
 & Saint Roman         &     547 M &  RTX 3090 &        77.1  &    34.6  & 111.7    &    4\,899           &       43\,570 &          13     &   $\times$ 390      \\
 \rot{\rlap{~first-come}}
 & CA21\_Bunds         &     976 M &  RTX A6000    &       134.5  &    60.9  & 195.4    &    4\,994           &       85\,651 &          11     &   $\times$ 438      \\
\hline 
 & Retz                &     145 M &  RTX 3090 &        19.9  &    46.7  &  66.7    &    2\,181           &       13\,130 &          11     &    $\times$ 196      \\
 & Palmyra             &     258 M &  RTX 3090 &        36.4  &    86.3  & 122.7    &    2\,110           &       21\,269 &          12     &    $\times$ 173      \\
 & Saint Roman         &     547 M &  RTX 3090 &        75.6  &   173.4  & 248.9    &    2\,198           &       43\,570 &          13     &    $\times$ 175      \\
 \rot{\rlap{~average}}
 & CA21\_Bunds         &     976 M &  RTX A6000 &       133.1  &   227.3  & 360.5    &    2\,707           &       85\,651  &         11     &    $\times$ 238       \\
\hline 
 & Retz                &     145 M &  RTX 3090 &        20.7  &   123.8  & 144.5    &    1\,006           &       13\,130 &          11     &    $\times$ 90      \\
 & Palmyra             &     258 M &  RTX 3090 &        37.3  &   231.6  & 268.9    &       962           &       21\,269 &          12     &    $\times$ 79      \\
 & Saint Roman         &     547 M &  RTX 3090 &        75.2  &   477.7  & 552.9    &       989           &       43\,570 &          13     &    $\times$ 79      \\
 \rot{\rlap{~weighted}}
 & CA21\_Bunds         &     976 M &  RTX A6000 &       133.8  &   638.5  & 772.3    &    1\,263           &       85\,651 &          11     &    $\times$ 111     \\
\hline 
\end{tabular*}
\caption{LOD construction times of our CUDA approach compared to Potree~\cite{SCHUETZ-2020-MPC}. Four sampling strategies were implemented and evaluated in CUDA, and compared to Potree's fastest \emph{random} sampling strategy.
All timings in milliseconds, throughput in million points per second (MP/s). }
\label{tab:our_performance}
\end{table*}

Table~\ref{tab:our_performance} compares LOD construction performance of our CUDA-based method to the CPU-based method of Potree. Potree offers a blue-noise sampling strategy that ensures a minimum distance between points at lower LODs, and a random sampling strategy that picks one random point per grid cell. The random sampling strategies of Potree and our CUDA approach are largely equivalent and therefore comparable. Since neither the blue-noise, nor the random sampling strategy of Potree compute color-filtered values, and because the blue-noise samples are not directly comparable to the voxelized samples of our CUDA approach, we always compare performances to Potree's fast random-sampling strategy. Furthermore, Potree is designed as an out-of-core application that simultaneously reads point data from disk, processes already loaded data, and writes result to disk. To increase the fairness of comparing processing times of our in-core implementation to Potree's out-of-core approach, we run Potree from a RAM-disk to ensure that disk I/O is not the limiting factor of its LOD construction process. We only had a mid-range CPU available for benchmarking Potree. Online benchmark repositories indicate that a high-end CPU with a similar price as our RTX 3090 -- the \emph{AMD Ryzen Threadripper 3960X} -- is about three times faster as our \emph{AMD Ryzen 7 2700X}~\cite{CPUBench}. 

The results in Table~\ref{tab:our_performance} indicate that GPGPU-based LOD construction approaches can be about \textbf{$\times$200 to $\times$300 times faster} compared to CPU-based approaches, with comparable quality (random). We also found that a first-come strategy further improves performance to about \textbf{$\times$400} with no apparent degradation of quality for our test data sets, but with potential quality reduction in scan-wise ordered terrestrial laser scans, as discussed in section~\ref{sec:quality}. Finally, our GPU-based method was able to construct LOD structures with higher quality \textbf{$\times$80 (weighted, in neighborhood) to $\times$200 (within a single cell) faster} than the fast random sampling strategy of Potree.

\subsection{Quality}
\label{sec:quality}

\begin{figure*}
    \centering
    \begin{subfigure}[t]{0.245\textwidth}
        \includegraphics[width=\textwidth]{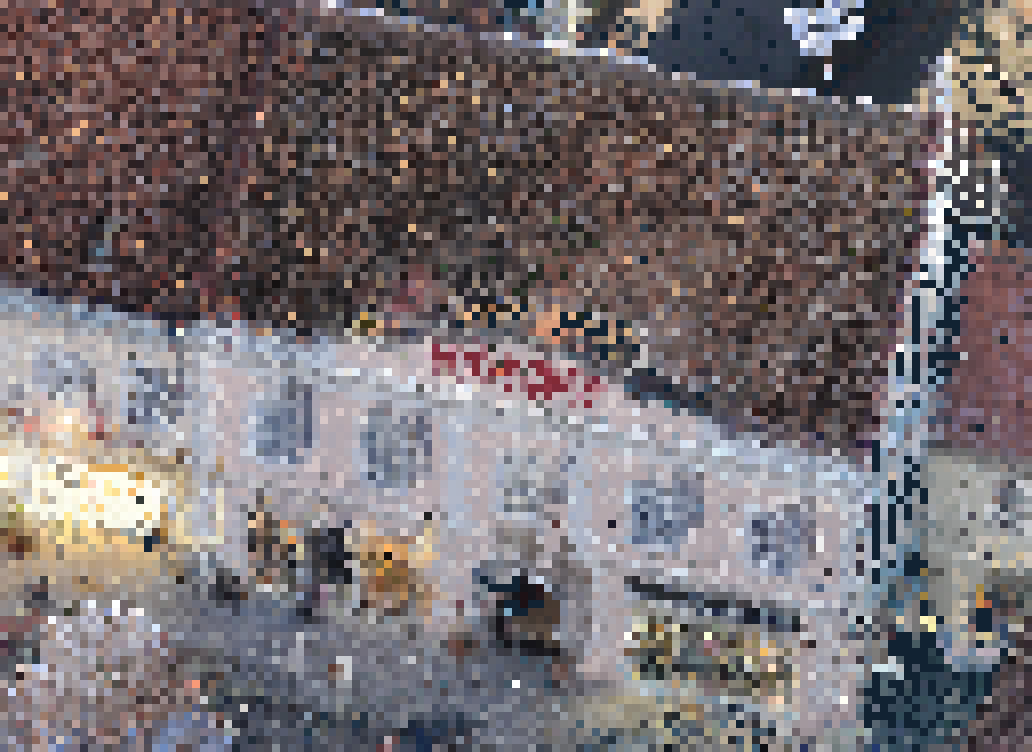}
    \end{subfigure}
    \hfill
    \begin{subfigure}[t]{0.245\textwidth}
        \includegraphics[width=\textwidth]{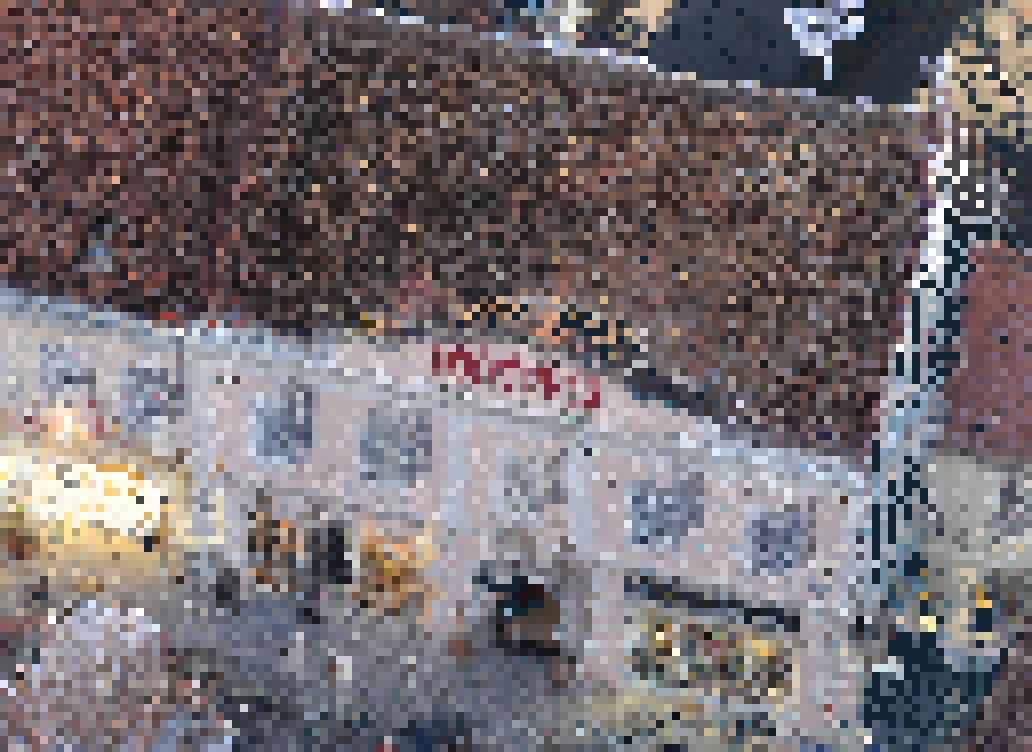}
    \end{subfigure}
    \hfill
    \begin{subfigure}[t]{0.245\textwidth}
        \includegraphics[width=\textwidth]{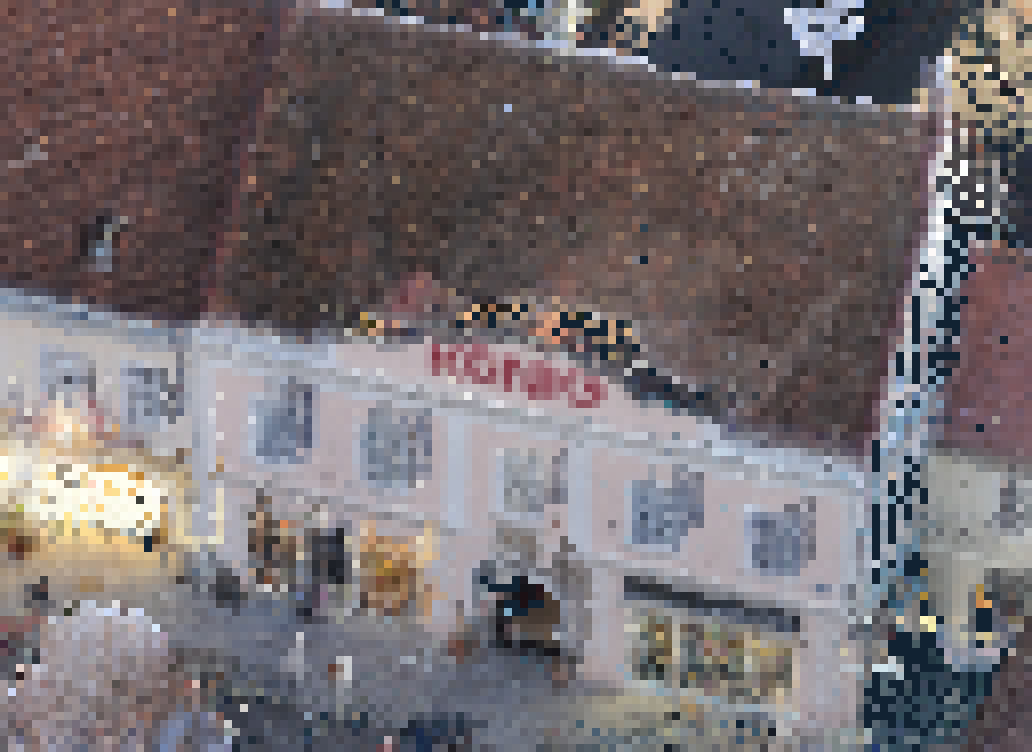}
    \end{subfigure}
    \hfill
    \begin{subfigure}[t]{0.245\textwidth}
        \includegraphics[width=\textwidth]{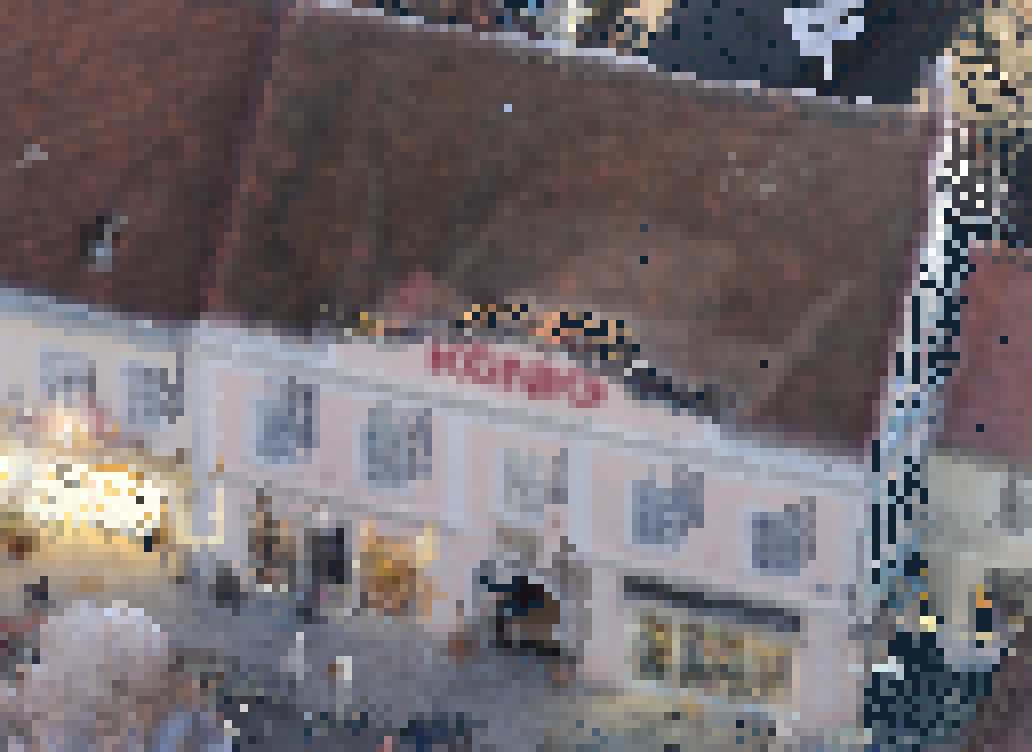}
    \end{subfigure}
    \begin{subfigure}[t]{0.245\textwidth}
        \includegraphics[width=\textwidth]{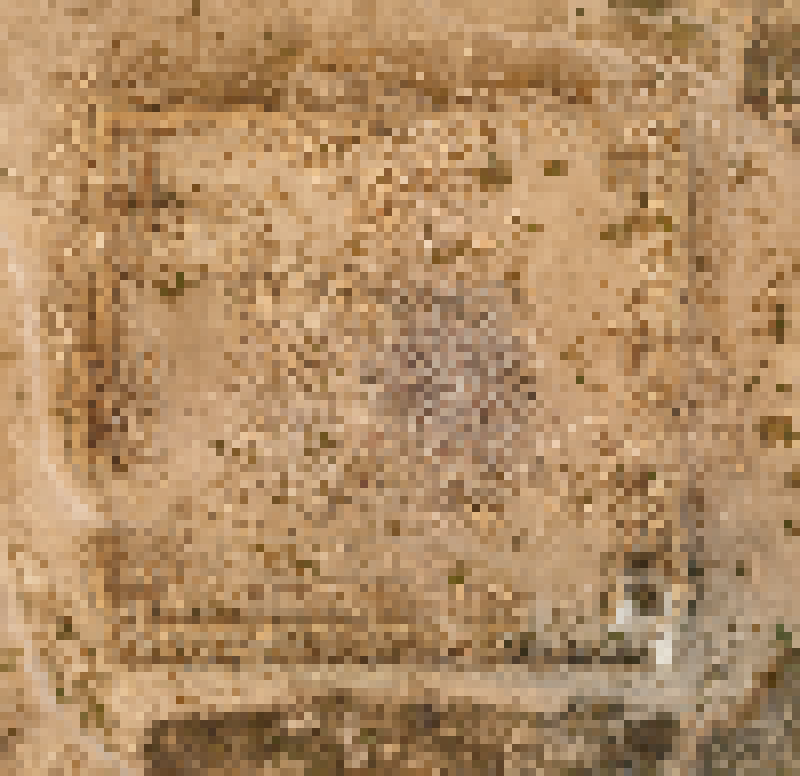}
    \end{subfigure}
    \hfill
    \begin{subfigure}[t]{0.245\textwidth}
        \includegraphics[width=\textwidth]{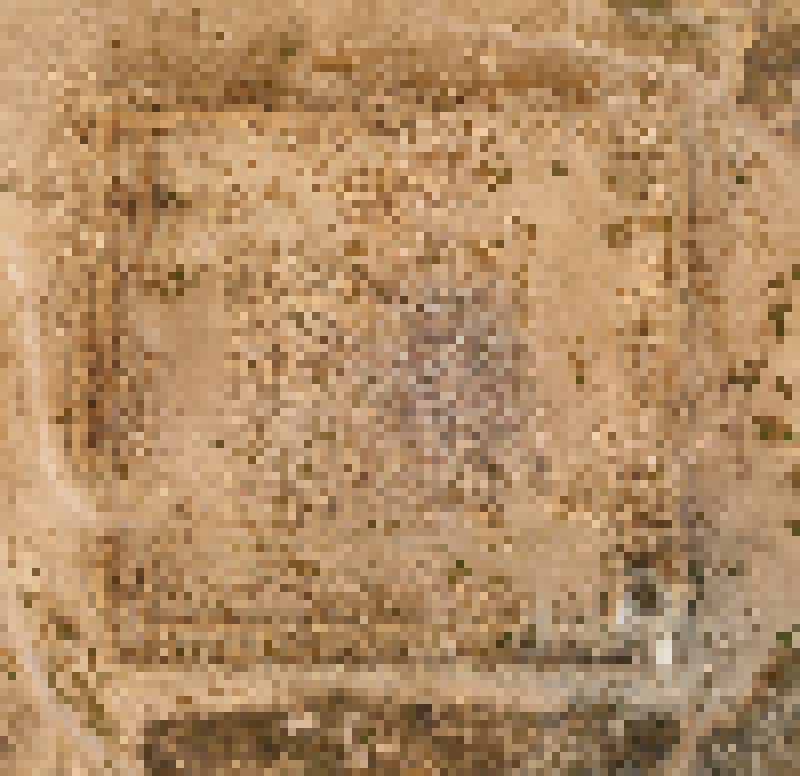}
    \end{subfigure}
    \hfill
    \begin{subfigure}[t]{0.245\textwidth}
        \includegraphics[width=\textwidth]{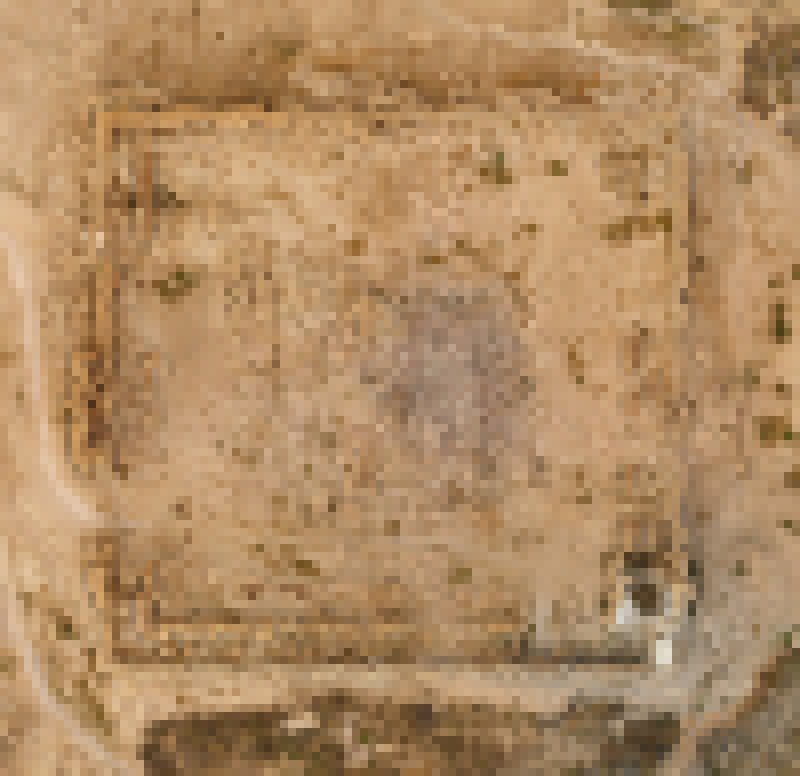}
    \end{subfigure}
    \hfill
    \begin{subfigure}[t]{0.245\textwidth}
        \includegraphics[width=\textwidth]{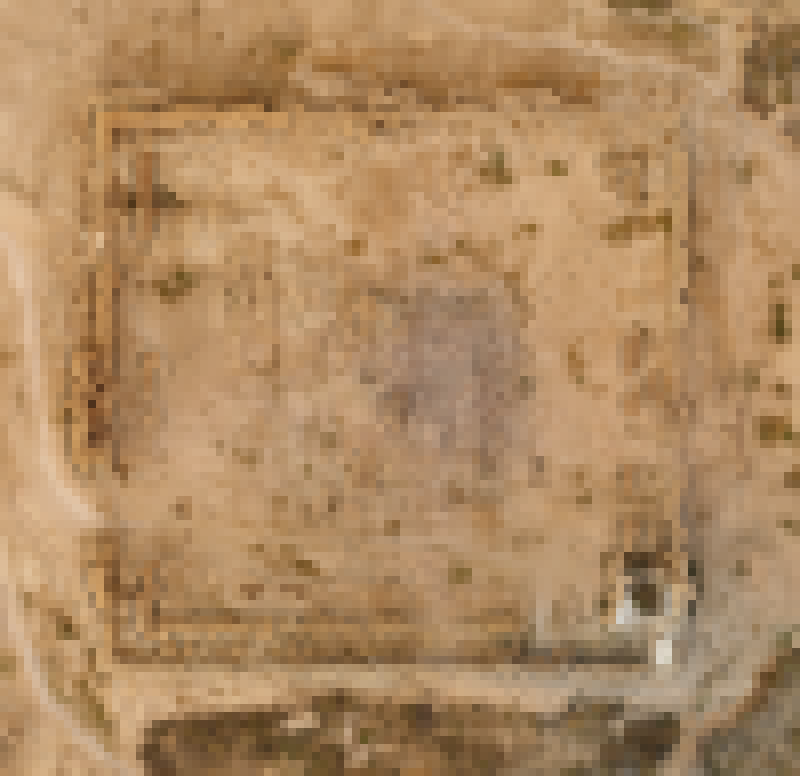}
    \end{subfigure}
    \begin{subfigure}[t]{0.245\textwidth}
        \includegraphics[width=\textwidth]{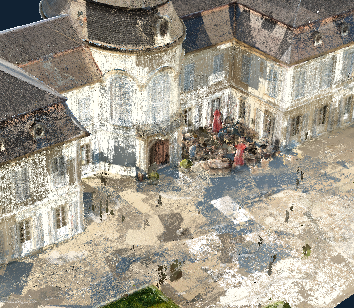}
        \caption{First-Come}
    \end{subfigure}
    \hfill
    \begin{subfigure}[t]{0.245\textwidth}
        \includegraphics[width=\textwidth]{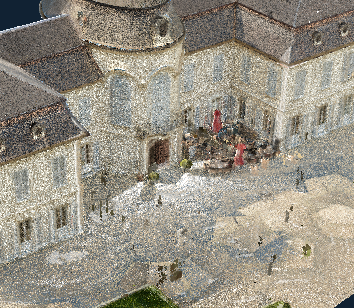}
        \caption{Random}
    \end{subfigure}
    \hfill
    \begin{subfigure}[t]{0.245\textwidth}
        \includegraphics[width=\textwidth]{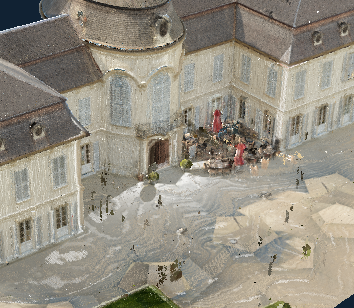}
        \caption{Average}
        \label{fig:quality_average}
    \end{subfigure}
    \hfill
    \begin{subfigure}[t]{0.245\textwidth}
        \includegraphics[width=\textwidth]{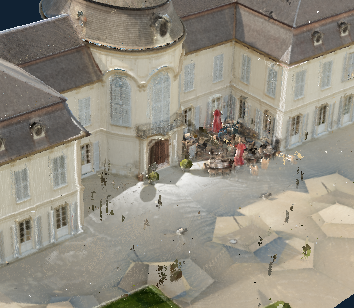}
        \caption{Weighted}
    \end{subfigure}
    \caption{Quality of the four sampling strategies. Differences between first-come and random are negligible in our performance test data sets, but we added \emph{Niederweiden} to the quality evaluation where it spectacularly fails. The average strategy significantly reduces noise and improves legibility of high-frequency features such as text, but still exhibits some noise/outliers due to sampling issues described in Figure~\ref{fig:sampling}. Weighted averages over voxel-boundaries addresses this issue and further improves quality.  }
	\label{fig:quality}
\end{figure*}

Figure~\ref{fig:quality} illustrates differences in quality between different sampling strategies. \emph{First-come} and \emph{random} sampling suffer from strong aliasing artifacts that manifest as noisy images that sparkle during motion -- similar to textured meshes without mip maps. In many data sets, both sampling strategies deliver similar results, but \emph{first-come} spectacularly fails for terrestrial laser scans where multiple scans with different exposure overlap. In that case, the random strategy appears better as it provides a uniform mix of all scans, rather than clusters of varying scans. \emph{Average} sampling within a single voxel provides significant improvements -- it reduces noise, makes high-frequency features such as text more readable, and it removes sparkling artifacts during motion. However, it still suffers from numerous outliers due to sampling issues shown in Figure~\ref{fig:sampling}. \emph{Weighted} average over adjacent cells removes these outliers and provides homogeneous, high-quality results similar to textured meshes with mip maps.

\section{Conclusion and Discussion}

We have shown that GPUs can accelerate the LOD construction of point clouds by two orders of magnitude compared to CPU-based approaches. There are, however, some limitations or potential issues that we would like to note and discuss.

\begin{itemize}
    \item \textbf{Out-of-core} processing was not evaluated -- the proposed approach is purely in-core, i.e., all data (input, temporary buffers, output), resides in GPU memory. However, we believe that this approach easily integrates in existing out-of-core and bottom-up LOD construction schemes~\cite{SCHUETZ-2020-MPC, Bormann:PCI} which first partition point clouds into large chunks (e.g., 10 million points). Each chunk could then be processed on the GPU by the method proposed in this paper.
    \item \textbf{Weighted-average color filtering} computes weighted averages within a neighborhood, but only for points and voxels within a node -- samples in close range but stored in another node are not considered. In practice, we found no perceivable issues with any of our test data sets, but sampling artifacts as shown in Figure~\ref{fig:quality_average} may occur, but rarely since they may occur between nodes instead of the $128^3$ voxels inside the node.
    \item \textbf{Optimal color filtering} was not evaluated -- we only showed that simple arithmetic averages within a cell or linear weighted samples within a range already significantly improve the quality. Better color filtering approaches are subject to future work.
    \item \textbf{View-dependent color filtering} may be an important topic for future work. Currently, surface samples that are visible from different directions collapse into a single voxel in lower levels of detail. For example, two sides of a wall may become a single voxel that holds only one color value for all view directions. Future work and implementations might revisit view-dependent voxels or impostors~\cite{FarVoxels, Wimmer-2001-Poi} or incorporate insights from neural radiance fields (NERFs), which evaluate the most suitable color value of a surface for each view direction~\cite{mildenhall2020nerf, KPLD21}, with spherical-harmonics-based approaches appearing particularly promising~\cite{yu_and_fridovichkeil2021plenoxels, yu2021plenoctrees}.
    \item \textbf{Voxels} are used in lower LODs mainly because they compress better than full-precision point coordinates, which is beneficial for streaming over web browsers. Their disadvantage is that measurement operations (distance, height, area, ...) will have reduced precision in lower levels of detail. However, with some modifications, our method supports full-precision point samples in lower LODs since in our implementation, voxels and points use the same struct with floating point coordinates. For the first-come and random sampling methods, we can simply deactivate the quantization to populate lower LODs with accurate point coordinates. For the average and weighted average methods, we could also create a list of accepted samples in addition to occupied voxels, and then use the average color values from the occupied voxels, and the coordinate value from the accepted samples to populate the lower LOD nodes.
\end{itemize}

The source code for this paper is available at \url{https://github.com/m-schuetz/CudaLOD}.

\section{Acknowledgements}

This research has been funded by FFG project LargeClouds2BIM. 

The authors wish to thank \emph{Iconem} for providing data sets \emph{Palmyra} and \emph{Saint Roman}; \emph{Schloss Schönbrunn Kultur- und Betriebs GmbH, Schloss Niederweiden} and \emph{Riegl Laser Measurement Systems} for providing the data set of Schloss Niederweiden; \emph{Riegl Laser Measurement Systems} for providing the data set of the town of \emph{Retz}; Bunds et al. and Open Topography for providing the data set CA21-Bunds~\cite{CA21Bunds}; and the \emph{Stanford University Computer Graphics Laboratory} for the \emph{Stanford Bunny}.

\printbibliography                

\end{document}